\documentclass[twocolumn]{aastex62}

\hypersetup{citecolor=blue}

\received{---}
\revised{---}
\accepted{---}
\submitjournal{ApJ}

\usepackage{ragged2e}
\usepackage[caption=false]{subfig}
\usepackage{booktabs}
\usepackage{soul}
\usepackage{amsmath}
\usepackage{cases}

\shorttitle{Reflection spectroscopy of GX 339--4 during its bright intermediate state transition}
\shortauthors{N Sridhar et al.}

\begin{document}

\title{\large{\textbf{Evolution of the accretion disk-corona during bright hard-to-soft state transition: \\A reflection spectroscopic study with GX 339--4}}}

\correspondingauthor{Navin Sridhar}
\email{navin.sridhar@columbia.edu}

\author[0000-0002-5519-9550]{Navin Sridhar}
\affil{Department of Astronomy, Columbia University, 550 W 120th St, New York, NY 10027, USA}

\author[0000-0003-3828-2448]{Javier A. Garc\'ia}
\affiliation{Cahill Center for Astronomy and Astrophysics, California Institute of Technology, Pasadena, CA 91125, USA}
\affiliation{Dr. Karl Remeis-Observatory and Erlangen Centre for Astroparticle Physics, Sternwartstr.~7, 96049 Bamberg, Germany}

\author{James F. Steiner}
\affiliation{MIT Kavli Institute for Astrophysics and Space Research, MIT, 70 Vassar Street, Cambridge, MA 02139, USA}

\author{Riley M. T. Connors}
\affiliation{Cahill Center for Astronomy and Astrophysics, California Institute of Technology, Pasadena, CA 91125, USA}

\author[0000-0003-2538-0188]{Victoria Grinberg}
\affiliation{Institute f\"{u}r Astronomie and Astrophysik (IAAT), Universit\"{a}t T\"{u}bingen, Sand 1, D-72076 T\"{u}bingen, Germany}

\author{Fiona A. Harrison}
\affiliation{Cahill Center for Astronomy and Astrophysics, California Institute of Technology, Pasadena, CA 91125, USA}



\begin{abstract}
We present the analysis of several observations of the black hole binary GX 339--4 during its bright intermediate states from two different outbursts (2002 and 2004), as observed by \textit{RXTE/PCA}. We perform a consistent study of its reflection spectrum by employing the \textsc{relxill} family of relativistic reflection models to probe the evolutionary properties of the accretion disk including the inner disk radius ($R_{\rm in}$), ionization parameter ($\xi$), temperatures of the inner disk ($T_{\rm in}$), corona ($kT_{\rm e}$), and its optical depth ($\tau$). Our analysis indicates that the disk inner edge approaches the inner-most stable circular orbit (ISCO) during the early onset of bright hard state, and that the truncation radius of the disk remains low ($\lesssim14 R_{\rm g}$) throughout the transition from hard to soft state. This suggests that the changes observed in the accretion disk properties during the state transition are driven by variation in accretion rate, and not necessarily due to changes in the inner disk's radius. We compare the aforementioned disk properties in two different outbursts with state transitions occurring at dissimilar luminosities, and find identical evolutionary trends in the disk properties, with differences only seen in corona's $kT_{\rm e}$ and $\tau$. We also perform an analysis by employing a self-consistent Comptonized accretion disk model accounting for the scatter of disk photons by the corona, and measure low inner disk truncation radius across the bright intermediate states, using the temperature dependent values of spectral hardening factor, thereby independently confirming our results from the reflection spectrum analysis.
\end{abstract}


\section{Introduction} \label{sec:intro}


One of the unsettled issues pertaining to black hole X-ray binaries (BHXB) is the origin of state transitions \citep{1972ApJ...177L...5T} seen during outbursts, and the associated evolution of various intrinsic physical parameters of the accretion disk \citep{2005Ap&SS.300..107H, 2006ARA&A..44...49R, 2012Sci...337..540F}. These state transitions are accompanied by changes in the overall luminosity of the source, relative flux of thermal and non-thermal emission, shape of the photon energy spectrum, onset of timing variability features in the light curves \citep[Quasi Periodic Oscillations,][]{1989ARA&A..27..517V}, and possible launching of ballistic radio jets and collimated outflows \citep{1999ARA&A..37..409M, 2004MNRAS.355.1105F, 2009MNRAS.396.1370F}. Typically, a given source will spend longer periods between outbursts in quiescence, i.e., in a state where little activity is seen accross the electromagnetic spectrum. The flux of the source is dominated by hard, non-thermal X-ray photons during the early onset as well as the late-time fading of an outburst, and constitute what is called the hard state (with photon index, $\Gamma\lesssim1.8$). The non-thermal X-ray emission is presumed to be arising from unsaturated Comptonization of disk photons by a cloud of Maxwellian electrons \citep[called `Corona';][]{1979Natur.279..506S}. In between these two hard states lie the soft and the intermediate states. The soft state, usually observed during the mid-outburst period has its spectra well described by thermal emission from the disk ($\sim 1$ keV), and occasionally by an additional power-law component with $\Gamma \gtrsim 2.5$. The more complicated intermediate states are characterized by a relative steepening of the non-thermal power-law component compared to the hard state, appearance of a thermal disk emission component, marked changes to the characteristic frequencies and fractional RMS (root mean square) of the power density spectrum, and launching of relativistic jets \citep[refer][for more details on intermediate states, and its sub-classifications into hard and soft-intermediate states]{2005Ap&SS.300..107H}. Comprehensive studies of stellar BHXB are therefore critical to deepen our understanding of black hole disks, coronae, jets, their relationships and evolution during the state transitions. From the HEASARC \textit{RXTE} archive, we have chosen GX 339--4 as our source for probing the disk and coronal evolutionary behavior, as this source is highly dynamic, and has undergone more than a dozen outbursts since its discovery \citep{2004MNRAS.355.1105F}.


GX 339--4 is an archetypal BHXB, whose first outburst was reported in 1973, and has been extensively studied since its discovery \citep[e.g.,][]{1973ApJ...184L..67M, 1986ApJ...308..635M, 1991ApJ...383..784M, 1997ApJ...479..926M, 2003A&A...400.1007C, 2003ApJ...583L..95H, 2004MNRAS.351..791Z, 2004MNRAS.347L..52G, 2005A&A...440..207B, 2015A&A...573A.120P}. It is known to be a part of a binary system located at a distance (d) of 5 kpc $<$ d $<$ 15 kpc \citep{2004ApJ...609..317H, 2004MNRAS.351..791Z, 2016ApJ...821L...6P}, with an orbital period of $\sim1.7$ days \citep{2003ApJ...583L..95H}, and whose inclination is not well known \citep{2002AJ....123.1741C, 2019MNRAS.488.1026Z}. A relatively recent dynamical study of the binary by \cite{2017ApJ...846..132H} has placed a constraint on the mass of GX 339--4 between $2.3M_{\odot}-9.5M_{\odot}$. This uncertainty stems from the lack of knowledge of the system's inclination precisely, which need not be the same as the inclination of the accretion disk. The inclination of the accretion disk, on the other hand, has been found to have an intermediate value ($30^\circ-60^\circ$) via different methods, namely X-ray reflection studies \citep{2015ApJ...808..122F, 2015ApJ...813...84G, 2016MNRAS.458.2199B, 2016ApJ...821L...6P}, and based on the separation of the double peaked emission lines \citep{2001MNRAS.320..177W}. This source has also exhibited ballistic jets observed in the radio band, which are typically associated with the intermediate or steep power-law (SPL) states \citep{1991ApJ...374..741M, 2004MNRAS.347L..52G, 2013ApJ...762..104S}. 

\begin{figure}[ht!]\centering 
\includegraphics[width=\linewidth]{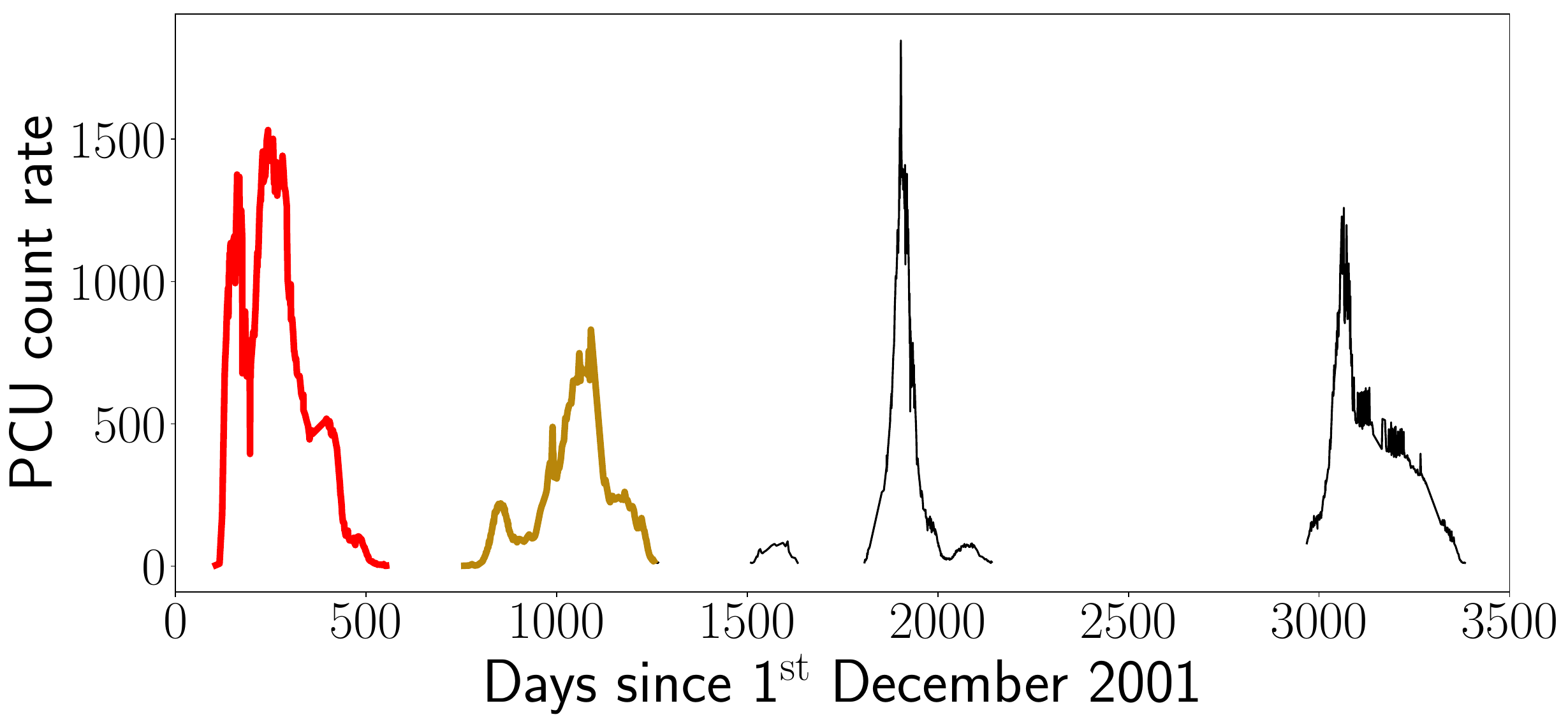}
\caption{\textit{RXTE}/PCA light curve of GX 339--4 showing its 4 prominent outbursts over a period of $\sim$10 years. The vertical axis shows the count rate (intensity), and plotted on the horizontal axis is the days that have passed since 1$^{\rm st}$ December 2001 (MJD$\sim$52244). The outbursts highlighted in red and golden-brown colors are the ones considered for the analysis presented in this paper. We call the red one as the `2002--2003 outburst' and the golden-brown one as the `2004--2005 outburst'.}
\label{fig:lc}
\end{figure}


\cite{2015ApJ...813...84G} perform the X-ray reflection spectroscopic analysis of the hard state spectra of GX 339--4 from the \textit{RXTE} archive, spanning a wide range of luminosities---1.6\% to 17\% of Eddington Luminosity ($L_{\rm Edd}$). With an assumed black hole spin of $a_{\star}=0.998$, they find the inner edge of the disk reaching $\sim$2 $R_{\rm g}$ (gravitational radius, $R_{\rm g}$ is given by $GM/c^2$, where M is the black hole's mass), as the luminosity increases from 1.6\% $L_{\rm Edd}$ to 17\% $L_{\rm Edd}$, and for the assumption of the disk at the innermost stable circular orbit (ISCO), they also constrain the spin of the source at $a_{\star}=0.95_{-0.05}^{+0.03}$. We choose to study the bright intermediate states of this source becauesm, it would be a natural segue from the hard state work of \cite{2015ApJ...813...84G} into the softer states, and also for the following reasons. Evidence for the intermediate states was reported for the first time by \cite{1997ApJ...479..926M} using the \textit{EXOSAT} observation of the source. During the intermediate states, the spectral hardness has an intermediate value---larger than soft state's, and smaller than hard state's spectral hardness. What makes it particularly interesting is the possible non-monotonicity in the evolution of parameters governing the bright intermediate state transition, and studying it may provide clues on how sources in general make transitions between hard and soft states. 

Of all the physical parameters of GX 339--4, the one that is highly debated is the physical evolution of the truncated accretion disk's inner radius ($R_{\rm in}$) (Refer Appendix A of \citealt{2016MNRAS.458.2199B} for a comparison of different $R_{\rm in}$ estimates), especially during the hard state. Based on \textit{Suzaku} data of GX 339--4, \cite{2012ApJ...753...65T} had reported that the optically thick disk is truncated significantly away from the ISCO during the hard-intermediate state. During the hard state, \cite{2000A&A...361..175M} had predicted that thermal conduction of heat from the corona will cause the inner disk to evaporate, leading to an optically thick and geometrically thin disk, that is truncated at some significant distance away from the black hole. While there is evidence for truncation during the hard state \citep{2019MNRAS.486.2137M}, and at luminosities below 0.1\% of the Eddington limit ($L_\mathrm{Edd}$) for GX 339--4 \citep{2009ApJ...707L..87T}, measurements of the reflection component in the bright hard state ($>5\%$ $L_\mathrm{Edd}$) have lead to estimates of inner radii very close to the ISCO for several sources, including GX 339--4 \citep{2015ApJ...813...84G, 2017ApJ...836..119S, 2018ApJ...855...61W, 2019arXiv190800965G}. While there exists abundant evidence for truncated disk at low/hard state, and the disk extending until ISCO by the time the source reaches bright soft state \citep{2004MNRAS.347..885G, 2010ApJ...718L.117S, 2010MNRAS.408..752P, 2012MNRAS.424.2504Z}, the evolution of the disk and the coronal parameters, especially the inner disk's radius during the bright intermediate states is not well understood. 


The reflection component of the X-ray spectrum results from the reprocessing of the Compton up-scattered non-thermal X-ray photons in the optically thick accretion disk. This reflected radiation  is adorned with several prominent features including the recombination continuum, fluorescent emission lines and absorption edges. These features provide information on the physical state of the matter and its composition in the strong gravitational regime near the black hole, and serves as a robust means of studying conditions near the event horizon. In particular, modeling the distortion in the prominently seen Fe K fluorescent line due to Doppler effects, gravitational red shift and light bending can contribute to the understanding of the spin of the black hole, the accretion disk's inner edge radius, and its inclination. Particularly, having spin fixed to a constant value of Thorne maximum \citep[$a_{\star}=0.998$;][]{1974ApJ...191..507T}, one can estimate the minimum truncation of the inner radius of the disk with much better sensitivity (Refer to \citealt{1972ApJ...178..347B} for a description of particle orbits in Kerr metric, and the relation between black hole's spin and the ISCO).

\begin{figure}[ht]
\includegraphics[width=1\linewidth]{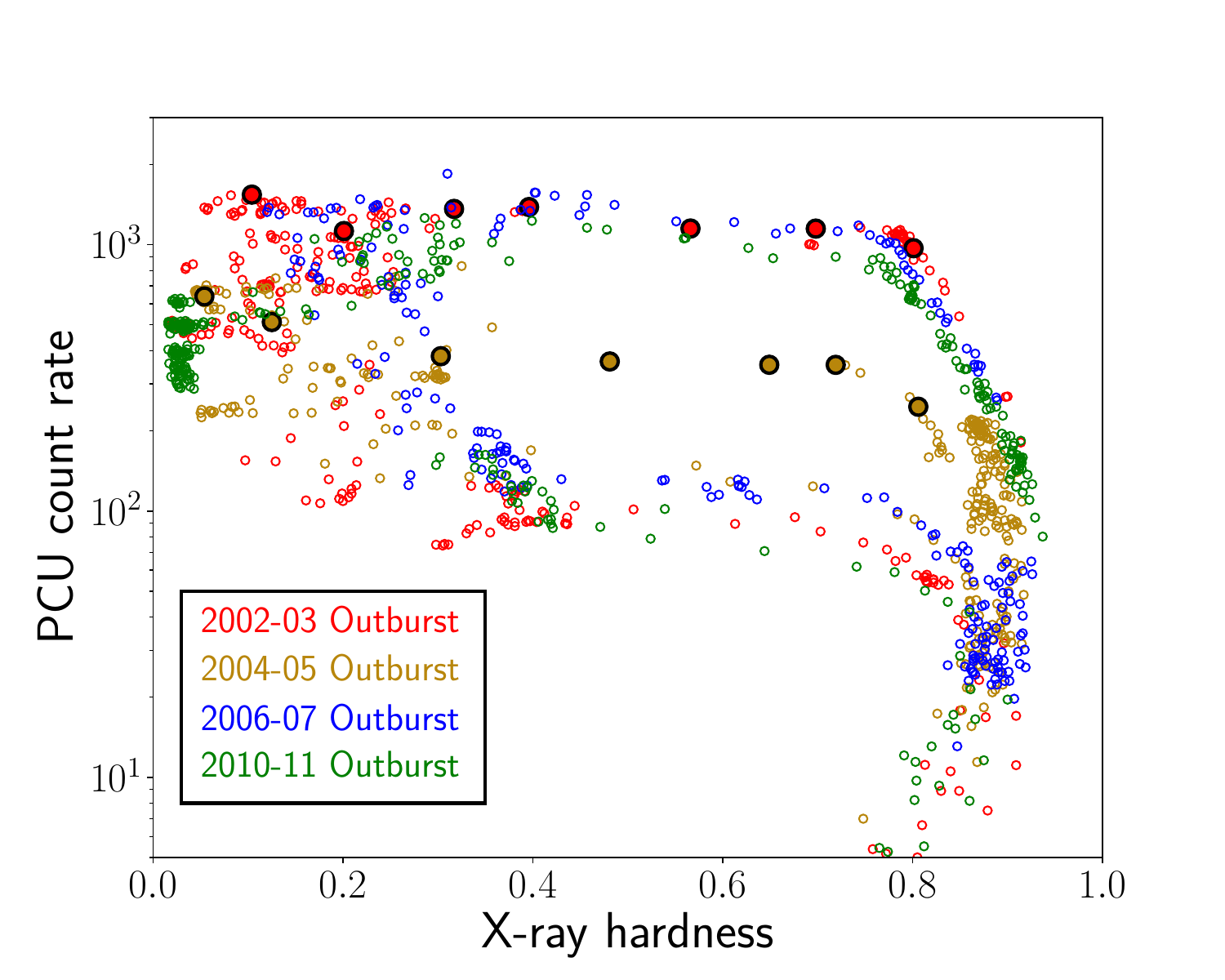}
\caption{Hardness-Intensity diagram (HID) from all the \textit{RXTE}/PCA observations of GX 339--4 spanning $\sim$10 years. Plotted on the horizontal axis is the X-ray hardness, defined as the ratio of the source counts in 8.6--18 keV to the counts in 5.0--8.6 keV energy band, and the vertical axis represents the PCU count rate (intensity). In this paper, we consider the 2002--2003 outburst (red) and the 2004--2005 outburst (golden-brown) for our analysis. Specifically, the interspaced highlighted red and golden-brown circles represent the positions of the observations on the HID, that are considered for the reflection analysis.}
\label{fig:HID}
\end{figure}

Although a study of GX 339--4 during the bright hard-to-soft state transition was performed by \cite{2008MNRAS.390..227D}, a probe of the reflection spectrum during this state transition remains relatively under-explored. Using a version of the relativistic reflection model (\texttt{relcillCp}, \citealt{2014ApJ...782...76G, 2014MNRAS.444L.100D}) that includes a physical Comptonization kernel (\texttt{nthComp}, \citealt{1996MNRAS.283..193Z}), we derive constraints on the evolution of system's parameters like the inner radius of the disk, its temperature, the disk ionization state, and the temperature and optical depth of the corona during the transit of the source from the bright hard to soft states of two outbursts with different transition luminosities (the 2002--2003 outburst with higher flux and the 2004--2005 outburst with lower flux), as observed by \textit{RXTE} (refer \citealt{2014MNRAS.442.1767P} for a reflection study of the source across various state transitions using different model and assumptions from ours). We also perform an analysis of the continuum spectra from the bright intermediate state transition by employing a self-consistent Comptonized accretion disk model. This model accounts for the scattering of the disk photons by the corona, to arrive at the disk and coronal properties independent of the reflection modelling, and to compare the results from one method against the other.

\begin{table*}[ht!] 
\centering
\caption{The \textit{RXTE}/PCA observation log of GX 339--4 across the bright intermediate states of the 2002--2003 and 2004--2005 outbursts (highlighted red and golden-brown circles in Figure \ref{fig:HID}), considered for the reflection analysis.}
\label{tab:observation_log}
\begin{tabular}{ c c c c c c }
  \hline
  \hline
  Outburst & ObsId & Start time & Exposure (s) & Counts ($\times 10^5$)&$HR$\\
  \hline
   & 40031-03-01-00 & 2002-04-18 02:42:59  & 1856 & 17.72 & 0.80\\
   & 70109-01-06-00 & 2002-05-06 20:12:04 & 1920 & 21.38 & 0.70\\  
   & 70110-01-10-00 & 2002-05-08 11:58:05 & 976 & 10.53 & 0.56\\  
   2002--2003 & 70109-04-01-00 & 2002-05-11 14:18:53 & 2720 & 33.74 & 0.40\\
   & 70110-01-12-00 & 2002-05-16 12:50:26 & 1248 & 14.94 &  0.30\\
   & 40031-03-03-04 & 2002-05-20 08:42:45 & 1632 & 15.69 & 0.20\\  
   & 70110-01-33-00 & 2002-07-28 19:54:20 & 880 & 11.54 & 0.10\\
  \hline    
   & 60705-01-67-01 & 2004-07-29 07:09:25 & 1024 & 2.49 & 0.80\\
   & 90704-01-01-00 & 2004-08-09 10:50:28 & 3168 & 10.88 & 0.72\\
   & 60705-01-69-01 & 2004-08-11 23:55:57 & 848 & 2.88 & 0.65\\
   2004--2005 & 60705-01-70-00 & 2004-08-13 23:07:57 & 720 & 2.44 & 0.48\\
   & 90704-01-03-00 & 2004-08-18 10:39:48 & 2736 & 9.45 & 0.30\\
   & 60705-01-76-00 & 2004-09-24 15:58:17 & 656 & 3.02 & 0.13\\
   & 90118-01-10-01 & 2004-10-28 12:08:33 & 2752 & 15.81 & 0.05\\
  \hline
\end{tabular}\\
\justify{
\tablecomments{The first column indicates the period of the considered outburst; the second column lists the \textit{RXTE} observation id; the third column lists the observation start time (UTC); the fourth column lists the corresponding observation's exposure time (in second), and the fifth column is the hardness ratio, defined as the ratio of the source count rate in the 8.6--18.0 keV to the count rate in the 5.0--8.6 keV energy band.}
}
\end{table*}

This paper is organized as follows: In \S\ref{sec:observation}, we describe the observation, data selection and reduction procedure, in \S\ref{sec:results} we provide the detailed analysis procedure, results from our analysis of reflection and continuum spectra, and the corresponding statistical (MCMC) analysis. In \S\ref{sec:discussion} we discuss the broad implications of our results, and summarise the same in \S \ref{sec:summary}. A test of the potential of \textit{RXTE}/PCA and the \textsc{relxill} model to detect large inner disk truncation values is provided in the Appendix.

\section{Observations and Data Reduction} \label{sec:observation}

We select our sample of the bright-intermediate state observations of the Galactic X-ray binary GX 339--4 from the huge inventory of data that \textit{RXTE} \citep{1999NuPhS..69...12S} had amassed during its lifetime. Figure \ref{fig:lc} shows four distinct outbursts of GX 339--4 spanning $\sim$10 years, as observed by \textit{RXTE}/PCA. We define here the bright-intermediate state to be the one bridging the brightest hard and soft states, with hardness ratio ($HR$) values in the range $0.05 < HR < 0.8$, where $HR$ is the ratio of the source counts at 8.6--18.0 keV to 5.0--8.6 keV energy band. This region is seen in Figure \ref{fig:HID} as the top horizontal row of each outburst. Figure \ref{fig:HID} is called the Hardness Intensity Diagram (HID), where the X-ray hardness (denoted above as $HR$) is plotted against the observed count rate (intensity). In the HID, outbursts 2002--2003, 2006--07 and 2010--11 can be seen to trace overlapping tracks across the bright intermediate states. We choose the earlier outburst as a representative of the three of them. The 2004--2005 outburst clearly traces a different path across the bright-intermediate states with a much smaller luminosity compared to the other three outbursts. So, we also choose observations from the 2004--2005 outburst for our analysis, to test whether there are any significant differences seen in the disk parameters of this outburst from the rest. The 2002--2003 and the 2004--2005 outbursts are highlighted in the $\sim$10 year light curve of the source (Figure \ref{fig:lc}) with red and golden-brown colors respectively. The individual pointings from the bright-intermediate states of the 2002--2003 and 2004--2005 outbursts that are considered for the reflection fitting procedure are listed in Table \ref{tab:observation_log}. 

The Proportional Counting Array (PCA) \citep{2006ApJS..163..401J} on-board the \textit{RXTE} mission comprised of five nearly identical Proportional Counting Units (PCU). The data from all five PCUs are reduced and background subtracted following the procedures described in \cite{2006ApJ...652..518M}. Data are taken in standard-2 mode, which provides coverage of the PCA bandpass every 16 s with exposure times ranging from 300 s to 5000 s. Background spectra are derived using the \textsc{pcabackest} tool with \texttt{pca\_bkgd\_cmvle\_eMv20111129.mdl} as the background model. Response files are generated using the tool \textsc{pcarmf} (version 11.7) considering the energy-to-channel conversion table (version e05v04) as described in \cite{2012ApJ...757..159S}. Following \cite{2014ApJ...794...73G}, we apply the tool \textsc{pcacorr} to all the data. This tool corrects for small imperfections in the spectrum, resulting in an increased sensitivity of PCA to faint spectral features. It therefore allows us to consider a much lower systematic uncertainty of 0.1\%. Calibration of the PCUs have been found to be uncertain above 45 keV, and the residuals seen below 3 keV cannot be explained by any plausible spectral features. So, we ignore the spectrum belonging to channels 1-4 and above 45 keV, and include only the rest, for our analysis.

\section{Results} \label{sec:results}

All of the spectral data analysis and model fitting are performed using \texttt{heasoft-6.23} and \textsc{xspec} version 12.9.1q \citep{1996ASPC..101...17A}. All models considered for the analysis described herein include the Galactic absorption effect by implementing the \texttt{TBabs} model \citep{2000ApJ...542..914W} with the corresponding abundances and cross sections set according to the \cite{1989GeCoA..53..197A} and \cite{1996ApJ...465..487V} photoelectric cross sections. This model is parametrized by the hydrogen column density, which for this source, we fix at a value of $5.9\times10^{21}$ cm$^{-2}$. This value is consistent with the determined value for this source from an analysis of \textit{RXTE}/PCA data by \cite{2015ApJ...813...84G}. We start our analysis by modeling the spectra with simple continuum models aimed at visualizing the Fe K reflection feature. We then extend our analysis by modeling all the key reflection signatures in the spectrum (i.e., Fe K emission, Fe K edge and Compton hump) using full-fledged relativistic reflection models. In the next few sections, we describe the various exercises performed with the bright-intermediate state data of the source, and the corresponding results.

\subsection{Ratio plots:\\ Model dependency and search for line broadening}\label{subsec:ratio}
The aim of this piece of work is to check if the Fe K reflection feature from the disk at different $HR$ values have any dependency on the chosen model of comptonization continuum. To start with, we consider the selected observations from the bright intermediate states of the 2002--2003 outburst (see \S\ref{sec:observation} and Table \ref{tab:observation_log}), and model them with three simple semi-physical models:
\begin{enumerate}
\item \texttt{TBabs*smedge(diskbb+cutoffpl+Gaussian)}
\item \texttt{TBabs*smedge(diskbb+powerlaw+Gaussian)}
\item \texttt{TBabs*smedge(diskbb+nthComp+Gaussian)}
\end{enumerate}
where, the galactic absorption is modeled with {\tt TBabs}, the multi-colored disk blackbody emission is modeled with {\tt diskbb} \citep{1984PASJ...36..741M, 1986ApJ...308..635M}, the Fe K reflection feature is modeled with a {\tt Gaussian}, and the smeared absorption edge at $\sim 7.1$ keV is modeled with {\tt smedge} \citep[implemented by Frank Marshall]{1991PhDT........55E}. Each model is different from the others by the choice of the implementation for modeling the non-thermal emission component. Models 1 and 2 use the phenomenological cut-off power-law ({\tt cutoffpl}) and a simple {\tt powerlaw} models respectively. Whereas model 3 employs the physically motivated \texttt{nthComp} (developed by \citealt{1996MNRAS.283..193Z} and extended by \citealt{1999MNRAS.309..561Z}) model to describe the comptonized component. 

\begin{figure}[t] \centering
\includegraphics[width=1\linewidth]{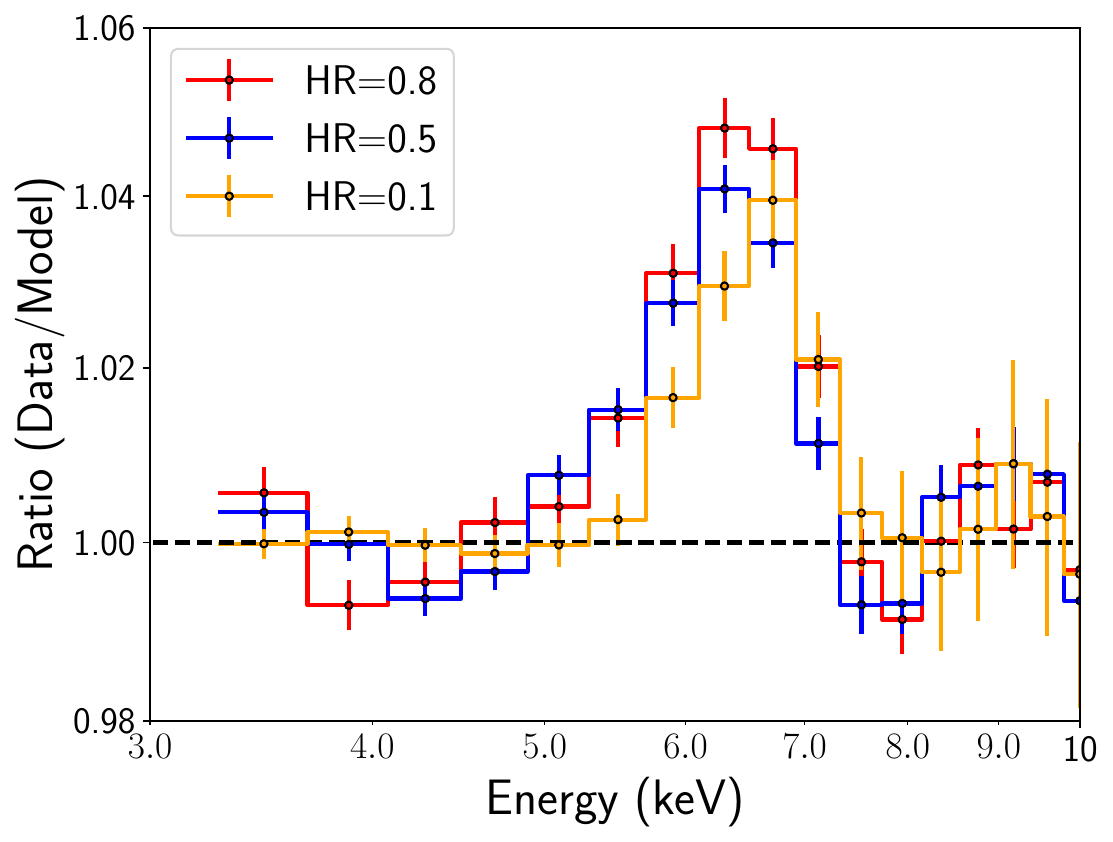}
\caption{Data/model ratio plots produced at different hardness values across the bright intermediate states of the 2002--2003 outburst of GX 339--4 using the model: \texttt{TBabs*smedge(diskbb+nthComp)}. No significant change in the width of the Fe K line is observed with decreasing spectral hardness.}
\label{fig:Fe_ratio}
\end{figure}

In all the three models, the \texttt{smedge} and \texttt{Gaussian} widths are frozen at 7 keV and 0.01 keV respectively, and the index for photoelectric cross-section in the \texttt{smedge} component is set to its standard value of -2.67. After fitting the spectra with these models, we retain the best fit parameters, and remove the respective {\tt Gaussian} components. These {\tt Gaussian}-less models are then used to visualize the Fe K reflection feature from the data/model ratio-plots. This approach validates that the profile of the Fe K reflection feature is insensitive to the model used to describe the Comptonized component of the spectrum. Reflection features during the intermediate states have also been observed by \cite{2014MNRAS.442.1767P}.

After establishing the insensitivity of the profile of Fe K reflection feature to different Comptonization models, we choose to model the spectrum with: \texttt{TBabs*smedge(diskbb+nthComp+Gaussian)}, and look out for any broadening in the reflection features, as we progress across a wide range of $HR$ values. We choose this model because, \texttt{nthComp}, unlike the \texttt{powerlaw} components, is a physically motivated model in describing the continuum shape from thermal Comptonization. Using this model, data/model ratio plots are recreated for observations in the bright hard state of the 2002--2003 outburst, following the procedures mentioned above. Figure \ref{fig:Fe_ratio} shows the ratio plot depicting the Fe K emission feature, plotted for three different representative values of $HR$ ($\simeq0.8, 0.5, 0.1$) across the bright intermediate states. Should the inner disk radius be significantly different between the soft and hard states, then the width of the iron line should also be different (see Appendix \ref{appendix:truncation}). However, no such significant change in broadening of the emission line can be seen.

\subsection{Fitting with the reflection model}\label{subsec:rf}

\begin{figure*}[ht]
\gridline{\fig{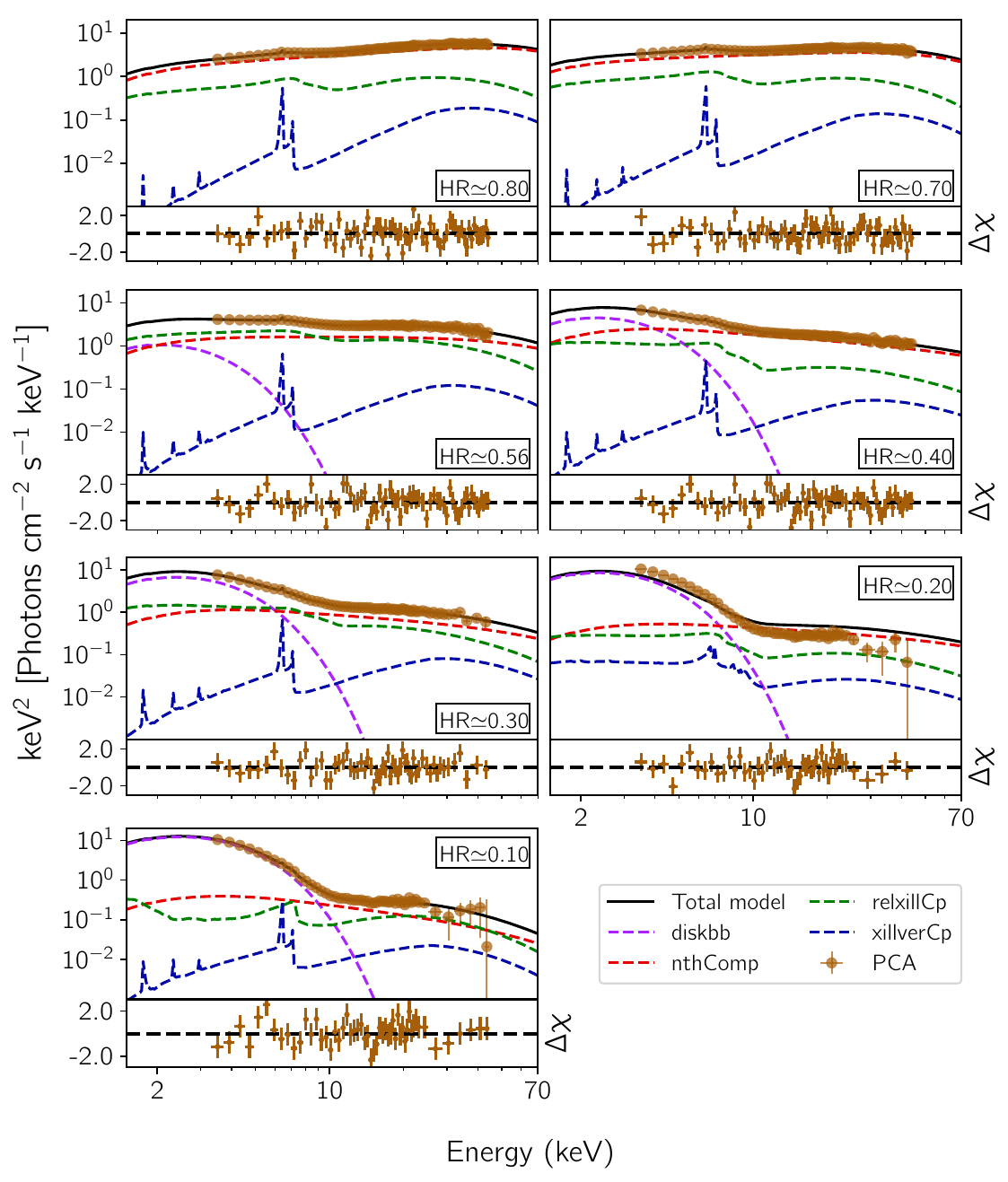}{0.5\textwidth}{(a) 2002--2003 Outburst}
			\fig{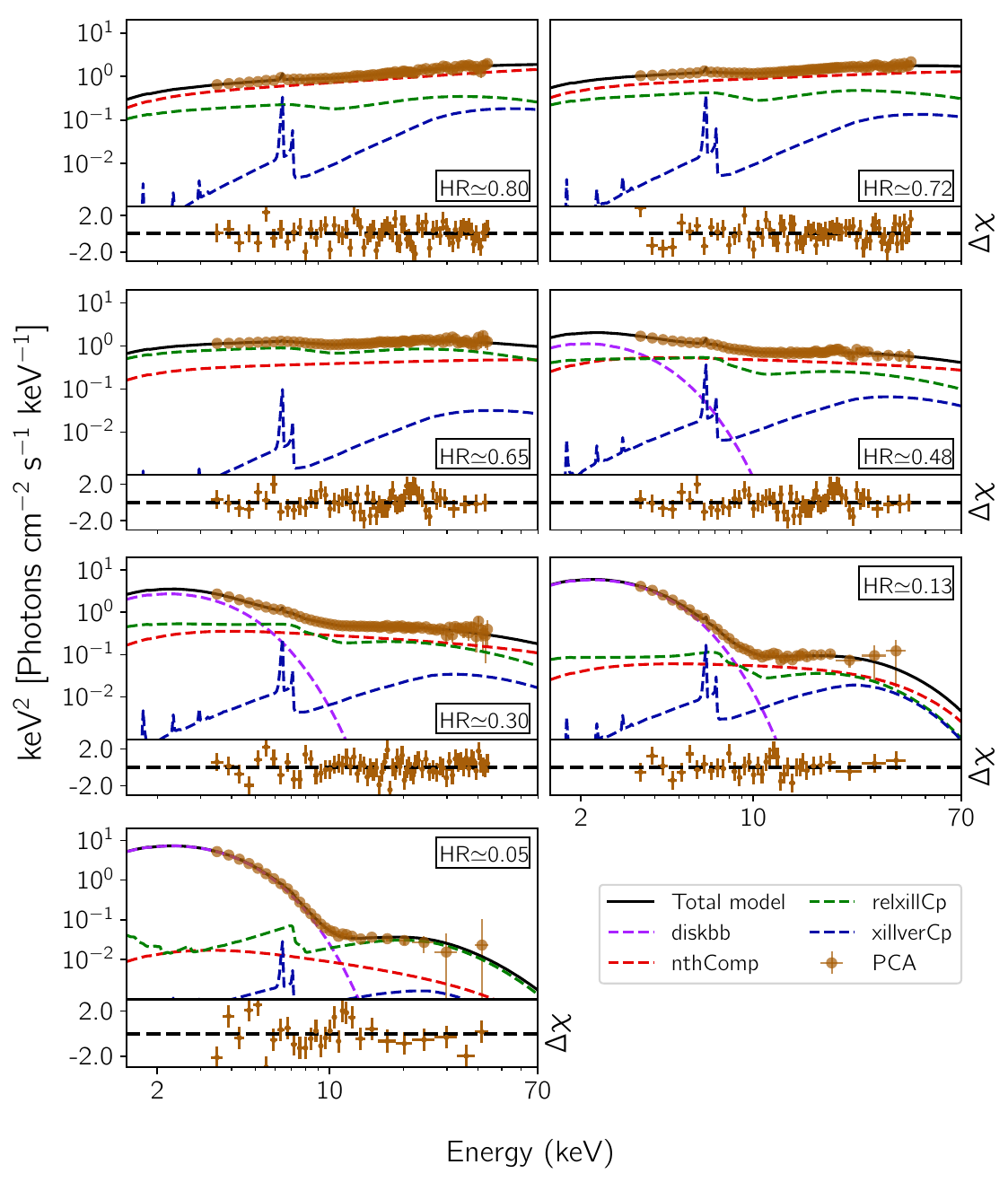}{0.5\textwidth}{(b) 2004--2005 Outburst}}
\caption{The extrapolated and unfolded \textit{RXTE}/PCA energy spectra of GX 339--4 across the bright intermediate states of the (a) 2002--2003 outburst (left segment) and (b) 2004--2005 outburst (right segment). The spectra are ordered according to their spectral hardness within an outburst, from the hardest spectrum at the top to the softest spectrum at the bottom. The top panel of the plot corresponding to each hardness ratio ($HR$) shows the \textit{RXTE}/PCA data (golden-brown circles) and the individual components (dashed lines) of the total fitted model (solid black line). The coronal emission is modeled with \texttt{nthComp} (red), disk emission is modeled with \texttt{diskbb} (purple), relativistic reflection component is modeled with \texttt{relxillCp} (green) and the distant (unblurred) reflection is modeled with {\tt xillverCp} (blue). The total model also includes a Galactic absorption component, modeled with \texttt{TBabs} (not shown here). The residuals from the fit ($\Delta \chi$) are plotted in the bottom panel of each spectral plot.}
\label{fig:spectra}
\end{figure*}

The spectrum of a typical black hole X-ray binary can be described well by the combination of a quasi-thermal blackbody emission due to the radiatively efficient accretion disk \citep{1995ApJ...445..780S}, a non-thermal power-law component due to Compton up-scattering of the thermal disk photons by a hot cloud of electron-positron pairs \citep[called Corona]{2004PThPS.155...99Z}, and the reflection component of the non-thermal emission by the optically thick accretion disk \citep{1993MNRAS.261...74R}, if seen. In our analysis, we model the thermal blackbody component of the spectrum using the \texttt{diskbb} model, and the non-thermal emission using the \texttt{nthComp} model. Following \cite{2015ApJ...813...84G}, we model the reflection features with a combination of relativistic disk reflection and a distant reflector for the narrow emission-line component.

The spectra of GX 339--4 has been shown to exhibit reflection features in its hard \citep{2015ApJ...813...84G} and soft states \citep{2004ApJ...606L.131M, 2008MNRAS.387.1489R}. In the previous section (\S\ref{subsec:ratio}), we showed the presence of the Fe K reflection feature across the bright intermediate states in the spectra. We account for that feature and other possible reflection components (namely, Fe K absorption edge, Compton hump) by modeling them with the physically rigorous \textsc{relxill} family of relativistic reflection models. The relativistically blurred reflection component is described using \texttt{relxillCp} \citep{2014ApJ...782...76G, 2014MNRAS.444L.100D}, and the narrow distant reflection from the disk is described using the \texttt{xillverCp} \citep{2010ApJ...718..695G, 2013ApJ...768..146G} model. The default shape of the illuminating continuum of the \textsc{relxill} and \textsc{xillver} models is a simple power-law with an exponential cut-off at higher energies. However in our model, we describe the non-thermal emission by the physically motivated \texttt{nthComp} component, and hence, the \texttt{relxillCp} and \texttt{xillverCp} flavors of the reflection models are used, which calculates the reflection spectrum by self-consistently using the \texttt{nthComp} continuum. The total model including Galactic absorption is given by: \texttt{TBabs(diskbb+nthComp+relxillCp+xillverCp)}.

In order to obtain better constraints of the key unknown parameters we are interested in (e.g., $R_\mathrm{in}$, $T_\mathrm{in}$, electron temperature $kT_\mathrm{e}$), we freeze the other parameters to their previously measured values. The hydrogen absorption column density is fixed at $N_{\rm H} = 5.9\times10^{21}$ cm$^{-2}$, the disk inclination is frozen at an intermediate value of $i = 45^{\circ}$, and the iron abundance $A_\mathrm{Fe}$ in the accretion disk at $A_\mathrm{Fe}=5$ (in Solar units). These numbers are consistent with the determined value of the corresponding parameters from the reflection analysis of \textit{RXTE}/PCA and \textit{NuSTAR} data of this source by \cite{2015ApJ...813...84G}, \cite{2015ApJ...808..122F} and \cite{2016ApJ...821L...6P}. These are the intrinsic parameters of the system, and are not expected to change over the course of outbursts. The spin value measurements of GX 339--4 by \cite{2004ApJ...606L.131M}, \cite{2008MNRAS.387.1489R}, \cite{2008ApJ...679L.113M}, \cite{2015ApJ...813...84G} and \cite{2016ApJ...821L...6P} have indicated that GX 339--4 is a black hole with a very high spin, close to the Kerr value \citep{1963PhRvL..11..237K}. Therefore, we fix the spin of the black hole at the limiting value of $a_{\ast}=0.998$, so that a wide range of $R_\mathrm{in}$ can be fully explored. One of the available parameters of the reflection models is the outer radius of the accretion disk ($R_\mathrm{out}$). We freeze it to a value large enough ($R_\mathrm{out} = 10^3~R_\mathrm{g}$) such that our measurements are not sensitive to its setting. This frozen value of $R_\mathrm{out}$ is much smaller than the actual radius of the outer disk, and only corresponds to that region of the disk which is relevant for X-ray reflection. The blurred and the unblurred reflection components (\texttt{relxillCp} and \texttt{xillverCp}) each accept a parameter for the reflection fraction (allowing for a thermal Comptonization continuum), which is defined as the ratio of the intensity emitted towards the disk, to that of the intensity escaping to infinity, in the frame of the primary source \citep[see][for more details]{2016A&A...590A..76D}. As we include an \texttt{nthComp} component explicitly accounting for the thermal Comptonization continuum, we freeze the reflection fraction parameter ($R_\mathrm{f}$) to -1, so that the {\tt relxillCp} component calculates only the reflected emission, and the overall model does not have two identical continuum components.

Some of the underlying assumptions of our analysis are as follows. Nearly neutral narrow Fe K lines have earlier been reported in bright Galactic binaries \citep{2015ApJ...808....9P}. We therefore assume that the unblurred, distant re-processing material is relatively cold and nearly neutral, and hence the log of the ionization parameter ($\log{\xi}$) corresponding to the \texttt{xillverCp} component is frozen at zero (where, the ionization parameter is given by $\xi = L/nR^2$, with $L$ being the ionizing luminosity, $n$ being the gas density and $R$ being the distance to the ionizing source). This assumption is also adopted by \cite{2015ApJ...813...84G} for this source.  For these fits, we also assume that the disk obeys the canonical emissivity profile, $\epsilon\propto r^{-q}$, where $q=3$ is the assumed value of emissivity index \citep{1989MNRAS.238..729F}. 

In addition to freezing some of the parameters to previously measured values, we also tie relevant parameters between different components of the model. For instance, the \texttt{nthComp} component accounts for the low energy roll-over due to the seed photons that originate from the accretion disk. Hence, we tie the \texttt{nthComp} input seed photon temperature ($kT_\mathrm{bb}$) to the inner disk temperature ($T_\mathrm{in}$ of the \texttt{diskbb} component). The previously mentioned assumption of a thermally Comptonized illuminating spectrum allows us to tie relevant parameters from the \texttt{nthComp} component like spectral index and electron temperature ($kT_\mathrm{e}$) with the ones in the \texttt{relxillCp} component. Additionally, the values for the $A_\mathrm{Fe}$, asymptotic power-law index $\Gamma$, disk inclination, and the input continuum in \texttt{xillverCp} are also linked with those of the broad reflection component ({\tt relxillCp}), as we find no empirical need to decouple these components' parameters.

With this setup, we simultaneously fit for the inner disk temperature ($T_\mathrm{in}$), photon index ($\Gamma$), electron temperature in the corona ($kT_\mathrm{e}$, with a hard upper limit at 400 keV), inner disk radius ($R_\mathrm{in}$), log of the disk ionization parameter ($\log{\xi}$), and the normalization values of the respective model components. The total model flux at different energy ranges is calculated using the \textit{flux} command in \textsc{xspec}, and the flux of the individual model components are evaluated by convolving the component, \texttt{cflux} (in \textsc{xspec} notation) with the required model components (e.g., with \texttt{relxillCp} and \texttt{nthComp}, used to calculate the reflection strength ($R_{\rm s}$), as mentioned in the upcoming sections). The spectral data, best fit model, underlying model components, and the corresponding residuals ($\Delta\chi$) from the best fit can be seen in Figure \ref{fig:spectra}. In the next couple of sections, we describe in detail the resultant best fit parameters attained for the considered observations by fitting with the model described above. The best fit parameters and the 90\% confidence intervals are obtained from the distributions of parameters, derived from performing the Markov Chain Monte Carlo analysis (see \S\ref{subsec:MCMC} for more details).

\subsubsection{The 2002--2003 outburst (Higher flux)}\label{subsubsec:2002--2003}

\begin{deluxetable*}{cccccccccc}[ht]

\tablecaption{Best fit parameters and the corresponding 90\% confidence intervals obtained from fitting the \textit{RXTE}/PCA spectra of GX 339--4 from its 2002--2003 outburst using the model: \texttt{TBabs(diskbb+nthComp+relxillCp+xillverCp)}, for different hardness ratio ($HR$) values across the bright intermediate states (see \S \ref{subsubsec:2002--2003}).} 
\tablewidth{0pt}
\tabletypesize{\scriptsize}

\tablehead{
\colhead{Spectral} &
\colhead{Parameters} &
\colhead{$HR\simeq0.8$} &
\colhead{$HR\simeq0.7$} &
\colhead{$HR\simeq0.56$} &
\colhead{$HR\simeq0.4$} &
\colhead{$HR\simeq0.3$} & 
\colhead{$HR\simeq0.2$} & 
\colhead{$HR\simeq0.1$} \\
\colhead{components} & 
}
\startdata
\texttt{TBabs} & {$N_{\rm H}$\tablenotemark{1} [cm$^{-2}$]} & \multicolumn{6}{c}{$5.9\times10^{21}$} \\
\texttt{relxillCp} & {$a_{\ast}$}\tablenotemark{2} & \multicolumn{6}{c}{0.998} \\
\texttt{relxillCp} & {$i$ [deg]} \tablenotemark{3}& \multicolumn{6}{c}{$45$} \\
\texttt{relxillCp} & {$A_{\rm Fe}$}\tablenotemark{4} [Solar]& \multicolumn{6}{c}{5} \\
\texttt{relxillCp} & { $R_{\rm f}$}\tablenotemark{5} & \multicolumn{6}{c}{-1} \\ 
\texttt{xillverCp} & {log$(\xi)$}\tablenotemark{6}  [erg cm s$^{-1}$]& \multicolumn{6}{c}{0} \\
\midrule
\texttt{diskbb} & $T_\mathrm{in}$\tablenotemark{7} [keV]& - & - & $0.61_{-0.09}^{+0.13}$ & $0.79_{-0.14}^{+0.15}$ & $0.82_{-0.01}^{+0.01}$ & $0.81_{-0.01}^{+0.01}$ & $0.87_{-0.01}^{+0.01}$ \\
\texttt{diskbb} & $N_{\rm disk}$\tablenotemark{8} ($10^3$) & $0$ & $0$ & $1.64_{-0.95}^{+1.95}$ & $2.14_{-0.13}^{+0.12}$ & $2.70_{-0.14}^{+0.12}$ & $3.63_{-0.25}^{+0.21}$ & $3.81_{-0.11}^{+0.09}$ \\
\texttt{nthComp} & $\Gamma$\tablenotemark{9} & $1.68_{-0.01}^{+0.01}$ & $1.83_{-0.01}^{+0.01}$ & $2.0_{-0.02}^{+0.03}$ & $2.41_{-0.02}^{+0.01}$ & $2.40_{-0.04}^{+0.03}$ & $2.36_{-0.09}^{+0.06}$ & $2.80_{-0.17}^{+0.07}$ \\
\texttt{nthComp} & $kT_\mathrm{e}$\tablenotemark{10} [keV]& $22.8_{-1.9}^{+2.7}$ & $19.6_{-1.4}^{+2.0}$ & $37.3_{-14.9}^{+311.0}$ & $246.0_{-142.6}^{+124.1}$ & $149.8_{-107.3}^{+194.7}$ & $96.2_{-74.9}^{+235.8}$ & $180.5_{-142.8}^{+170.9}$ \\
\texttt{nthComp} & $N_{\rm nthC}$\tablenotemark{11} & $0.96_{-0.18}^{+0.35}$ & $1.77_{-0.38}^{+0.28}$ & $0.89_{-0.28}^{+0.59}$ & $1.37_{-0.09}^{+0.13}$ & $0.70_{-0.19}^{+0.18}$ & $0.19_{-0.19}^{+0.07}$ & $0.23_{-0.05}^{+0.06}$ \\
\texttt{relxillCp} & $R_\mathrm{in}$\tablenotemark{12} $[R_\mathrm{ISCO}]$ & $6.85_{-2.47}^{+4.15}$ & $<4.39$ & $<4.44$ & $1.71_{-0.24}^{+0.50}$ & $<2.32$ & $<2.28$ & $1.85_{-0.69}^{+1.03}$ \\
\texttt{relxillCp} & log$(\xi)$\tablenotemark{13} [erg cm s$^{-1}$]& $3.82_{-0.20}^{+0.09}$ & $3.95_{-0.29}^{+0.10}$ & $4.29_{-0.57}^{+0.07}$ & $4.20_{-0.16}^{+0.12}$ & $4.35_{-0.21}^{+0.12}$ & $4.20_{-0.38}^{+0.24}$ & $1.86_{-0.69}^{+1.03}$ \\
\texttt{relxillCp} & $N_{\rm rel}$\tablenotemark{14} $(10^{-3})$ & $8.89_{-0.75}^{+0.09}$ & $11.8_{-1.0}^{+1.0}$ & $21.1_{-7.5}^{+3.3}$ & $21.3_{-2.4}^{+1.6}$ & $19.2_{-0.4}^{+4.7}$ & $7.0_{-2.4}^{+3.8}$ & $5.6_{-4.3}^{+4.3}$ \\
\texttt{xillverCp} & $N_{\rm xil}$\tablenotemark{15} $(10^{-3})$ & $3.43_{-0.65}^{+0.10}$ & $2.37_{-0.10}^{+0.13}$ & $<4.39$ & $5.0_{-1.2}^{+2.5}$ & $6.53_{-0.40}^{+0.21}$ & $1.6_{-1.6}^{+2.3}$ & $<16.95$ \\
\midrule
$2-10$ keV Flux\tablenotemark{16} & [10$^{-8}$erg cm$^{-2}$ s$^{-1}$] & 1.06 & 1.71 & 1.89 & 1.91 & 2.09 & 1.22 & 1.92\\
Luminosity & $L/L_\mathrm{Edd}~(\%)$\tablenotemark{17} & 21.1 & 29.7 & 34.61 & 45.8 & 46.4 & 23.9 & 31.2 \\
Reflection strength & $R_\mathrm{s}$\tablenotemark{18} & $0.23_{-0.05}^{+0.05}$ & $0.26_{-0.05}^{+0.04}$ & $0.66_{-0.39}^{+0.30}$ & $0.74_{-0.07}^{+0.07}$ & $0.53_{-0.28}^{+0.17}$ & $0.36_{-0.21}^{+0.11}$ & $1.32_{-0.64}^{+0.37}$\\
\midrule
& {$\chi^{2}_{\nu}$}\tablenotemark{19} & $\frac{72.49}{62}=1.17$ & $\frac{64}{63}=1.02$ & $\frac{57.43}{62}=0.93$ & $\frac{69.48}{62}=1.12$ & $\frac{51.18}{62}=0.83$ & $\frac{52.23}{62}=0.84$ & $\frac{53.24}{62}=0.86$ \\
[1ex]
\enddata
\tablecomments{$^{1}$ Hydrogen column density; $^{2}$ Dimensionless spin parameter of the black hole ($a_{\star}$ = $cJ/GM^2$, where $J$ is the angular momentum of the black hole); $^{3}$ Disk inclination; $^{4}$ Iron abundance of the material in the accretion disk; $^{5}$ Reflection fraction, defined as the ratio of the reflected flux to the flux in the power-law component, in 20--40 keV band; $^{6,13}$ $\log$ of the ionization parameter ($\xi$) of the accretion disk, where $\xi=L/nR^{2}$, with \textit{L} as the ionizing luminosity, \textit{n} as the gas density and \textit{R} as the distance to the ionizing source; $^{7}$ Temperature at the inner disk; $^{9}$ Asymptotic power-law photon index; $^{8}$ Normalization of {\tt diskbb}, given by $N_{\rm disk}=(R_{\rm in}/D_{10})^2\cos{\theta}$, where $R_{\rm in}$ is the apparent inner disk radius (in km), $D_{10}$ is the distance to the source (in units of 10 kpc), and $\theta$ is the angle of inclination of the disk (in deg) \citep{1998PASJ...50..667K}; $^{10}$ Electron temperature determining the high energy roll-over; $^{11}$ When the normalization of {\tt nthComp} is unity, the corresponding model flux is 1 photon keV$^{-1}$cm$^{-2}$s$^{-1}$ at 1 keV; $^{12}$ Inner radius of the accretion disk; $^{14,15}$ Normalization of {\tt xillverCp} ($N_{\rm xil}$) is defined such that for an incident spectrum with flux $F_X(E)$ incident on a disk with density $n_e$, and with ionization parameter $\xi$, the integral $\int_{0.1 keV}^{1MeV} F_X(E)dE$ is evaluated to be $10^{20}\frac{n_e\xi}{4\pi}$. For {\tt relxillCp}, the definition of the normalization ($N_{\rm rel}$) is identical to that of {\tt xillverCp}, except that the flux reaching the observer is modified due to relativistic effects \citep[see][]{Dauserb}; $^{16}$ Unabsorbed flux (For reference, 1 Crab = $2.2\times10^8$ erg cm$^{2}$s$^{-1}$ in 2--10 keV band); $^{17}$ Luminosity of the source in terms of its Eddington limit, computed assuming the mass of the black hole to be $M = 10M_{\odot}$, a distance of $D = 8$ kpc (corresponding to $L_\mathrm{Edd}=1.25\times10^{38}$ erg s$^{-1}$), and from the 0.01--100 keV fluxes derived from the best-fit parameters; $^{18}$ Reflection strength, defined to be the ratio of the observed fluxes of the reflected component ({\tt relxillCp}) and the incident component ({\tt nthComp}) in the 20--40 keV energy band; $^{19}$ Reduced $\chi^2$, defined as the ratio of the best fit $\chi^2$ to the number of degrees of freedom ($\nu$).}
\label{tab:2002--2003}
\end{deluxetable*}

The HID of GX 339--4 during 2002--2011 (see Figure \ref{fig:HID}) shows three outbursts (2002--2003, 2006-07 and 2010-11) tracing similar paths in the spectral hardness-luminosity space. Out of them, we choose the 2002--2003 outburst as a representative outburst of that group, and perform reflection analysis. We begin by modeling the spectra from the bright intermediate states of 2002--2003 outburst in a temporal sequence, starting with the hard state ($HR\simeq0.8$) spectrum and gradually proceeding towards the soft state ($HR\simeq0.1$) (see Table \ref{tab:observation_log} and red markers in Figure \ref{fig:HID}). The best fit parameters and the 90\% confidence intervals obtained by fitting the 2002--2003 observations with our reflection model are listed in Table \ref{tab:2002--2003}.

During the analysis, we notice that the hardest two spectra considered here ($HR\simeq0.8$ and 0.7) do not exhibit a significant thermal emission component, and including a {\tt diskbb} component with free parameters is not required. The absence of a distinct blackbody emission component in the spectra can be seen in the top two panels of Figure \ref{fig:spectra}a. Therefore, we freeze the normalization parameter of the {\tt diskbb} component to 0, and as a result we are unable to arrive at a good constraint on the temperature of the inner disk. As the outburst progresses towards the intermediate states ($HR\gtrsim0.56$ in our selection of observations), we see a need for the thermal blackbody component in order to describe the spectra well, and therefore include the {\tt diskbb} component, and fit for its normalization as a free parameter. In all the forthcoming fits with softer spectra, we observe the $T_{\rm in}$ and {\tt diskbb} norm parameters to gradually increase all the way across the bright intermediate states, reaching a value of $T_{\rm in}=0.87_{-0.01}^{+0.01}$ keV and norm $=3.81_{-0.11}^{+0.09}\times10^3$ respectively, by the time the spectral hardness reaches a value of $HR\simeq0.1$ (see eq. \ref{eq:norm} in \S \ref{subsubsec:Tin} for a discussion into {\tt diskbb} norm).

For the cases of $HR\simeq 0.8$ and 0.7, the asymptotic spectral indices are found to be $\Gamma = 1.68_{-0.01}^{+0.01}$ and $1.83_{-0.01}^{+0.01}$ respectively, which are typical of the hard state \citep{2005Ap&SS.300..107H, 2006ARA&A..44...49R}. As the spectrum gets softer, the non-thermal tail is seen to get steeper as expected (see last two panels of Figure \ref{fig:spectra}a), thus resulting in an increasing trend in $\Gamma$, as $HR$ decreases. This increase in $\Gamma$ culminates at a value of $2.80_{-0.17}^{+0.07}$ for the spectrum with $HR\simeq0.1$. From the fits, we constrain the coronal electron temperatures ($kT_{\rm e}$) to values between 18.2 keV and 25.5 keV during the bright hard to intermediate state transition ($0.8\geqslant HR\geqslant0.56$). However, the values of the electron temperature cannot be constrained well during the intermediate to soft state transition ($0.56\geqslant HR\geqslant0.1$). 

One of the key physical parameters that we are interested in understanding the evolution of, is the inner radius of the accretion disk, as the outburst evolves from the bright hard to the bright soft state. In the earlier section (see \S \ref{subsec:ratio}), we qualitatively demonstrated from the unchanging nature of the width of Fe K reflection profile that the disk truncation is relatively stable across the bright intermediate state transition. This result is verified by fitting the spectrum with the state-of-the-art reflection model that we employ. Fitting for the $R_{\rm in}$ parameter of the {\tt relxillCp} component yields a value of $R_{\rm in}=6.85_{-2.47}^{+4.15}~R_{\rm ISCO}$ ($5.4~R_{\rm g}-13.6~R_{\rm g}$) for the hardest spectrum in our sample ($HR\simeq0.8$). $R_{\rm in}$ then converges to $\lesssim 4.4~R_\mathrm{ISCO}$ ($\lesssim 5.4~R_{\rm g}$) by the time the spectral hardness reaches $HR\simeq0.7$, and remains close to ISCO all the way across the bright intermediate states, until the bright soft state ($HR \simeq 0.1$). Our results are incompatible with an inner radius placed at hundreds of $R_{\rm g}$, as has been reported elsewhere \citep{2009MNRAS.396.1415C, 2010MNRAS.407.2287D, 2012ApJ...753...65T, 2015A&A...573A.120P}.

As a part of our analysis, we quantify the reflection strength ($R_{\rm s}$) by computing the ratio of the observed fluxes of the reflected component (\texttt{relxillCp}) and the incident component ({\tt nthComp}) in the 20--40 keV energy band. The reflection strength is seen to be weakest ($R_{\rm s}=0.23^{+0.05}_{-0.05}$) during the bright hard state ($HR\simeq0.8$), and during the bright intermediate states ($0.7\lesssim HR\lesssim0.2$), $R_{\rm s}$ is seen to oscillate between 0.21 and 0.47. This is followed by a steep increase in its value, culminating at $R_{\rm s}=1.32_{-0.64}^{+0.37}$, by the time the system reaches the bright soft state ($HR\sim 0.1$). This is a $2.7
\sim 8.8$ fold rise in the reflection strength, from its value during the inception of bright intermediate state transition.

\subsubsection{The 2004--2005 outburst (Lower flux)}\label{subsubsec:2004--2005}
\begin{deluxetable*}{cccccccccc}[ht]
\tablecaption{Best fit parameters and the corresponding 90\% confidence intervals obtained from fitting the \textit{RXTE}/PCA spectra of GX 339--4 from its 2004--2005 outburst using the model: \texttt{TBabs(diskbb+nthComp+relxillCp+xillverCp)}, for different hardness ratio ($HR$) values across the bright intermediate states (see \S \ref{subsubsec:2002--2003}).} 
\tabletypesize{\scriptsize}
\tablewidth{0pt}

\tablehead{
\colhead{Spectral} &
\colhead{Parameters} &
\colhead{$HR\simeq0.8$} &
\colhead{$HR\simeq0.72$} &
\colhead{$HR\simeq0.65$} &
\colhead{$HR\simeq0.48$} &
\colhead{$HR\simeq0.3$} & 
\colhead{$HR\simeq0.13$} &
\colhead{$HR\simeq0.05$} \\
\colhead{components} & 
}
\startdata
\texttt{TBabs} & {$N_{\rm H}$\tablenotemark{1} [cm$^{-2}$]} & \multicolumn{7}{c}{$5.9\times10^{21}$} \\
\texttt{relxillCp} & {$a_{\ast}$}\tablenotemark{2} & \multicolumn{7}{c}{0.998} \\
\texttt{relxillCp} & {$i$ [deg]}\tablenotemark{3} & \multicolumn{7}{c}{$45$} \\
\texttt{relxillCp} & {$A_\mathrm{Fe}$\tablenotemark{4} [Solar]} & \multicolumn{7}{c}{5} \\
\texttt{relxillCp} & { $R_\mathrm{f}$}\tablenotemark{5} & \multicolumn{7}{c}{-1} \\ 
\texttt{xillverCp} & {log$(\xi)$}\tablenotemark{6} [erg cm s$^{-1}$]& \multicolumn{7}{c}{0} \\
\midrule
\texttt{diskbb} & $T_\mathrm{in}$\tablenotemark{7} [keV]& - & - & - & $0.70_{-0.04}^{+0.03}$ & $0.72_{-0.02}^{+0.01}$ & $0.75_{-0.02}^{+0.02}$ & $0.78_{-0.02}^{+0.02}$ \\	
\texttt{diskbb} & $N_{\rm disk}$\tablenotemark{8} ($10^3$) & $0$ & $0$ & $0$ & $0.81_{-0.18}^{+0.30}$ & $2.26_{-0.11}^{+0.10}$ & $3.48_{-1.4}^{+0.3}$ & $3.62_{-0.32}^{+0.33}$\\
\texttt{nthComp} & $\Gamma$\tablenotemark{9} & $1.75_{-0.12}^{+0.02}$ & $1.79_{-0.02}^{+0.01}$ & $2.03_{-0.16}^{+0.04}$ & $2.48_{-0.28}^{+0.04}$ & $2.38_{-0.04}^{+0.03}$ & $2.48_{-0.27}^{+0.42}$ & $2.72_{-0.36}^{+0.16}$\\
\texttt{nthComp} & $kT_\mathrm{e}$\tablenotemark{10} [keV]& $240.3_{-133.7}^{+131.3}$ & $307.9_{-105.3}^{+75.0}$ & $327.7_{-97.6}^{+57.5}$ & $289.0_{-139.1}^{+90.9}$ & $270.0_{-259.8}^{+106.2}$ & $152.3_{-142.3}^{+199.4}$ & $110.0_{-96.9}^{+227.4}$\\
\texttt{nthComp} & $N_{\rm nthC}$\tablenotemark{11} & $0.36_{-0.17}^{+0.03}$ & $0.48_{-0.05}^{+0.49}$ & $0.39_{-0.06}^{+0.61}$ & $0.33_{-0.11}^{+0.06}$ & $0.15_{-0.06}^{+0.03}$ & $0.03_{-0.01}^{+0.02}$ & $0.07_{-0.01}^{+0.01}$\\
\texttt{relxillCp} & $R_\mathrm{in}$\tablenotemark{12} $[R_\mathrm{ISCO}]$ & $2.24_{-0.94}^{+3.99}$ & $<1.51$ & $1.27_{-0.12}^{+0.56}$ & $<2.86$ & $<2.06$ & $<3.47$ & $<2.93$\\
\texttt{relxillCp} & $\log(\xi)$\tablenotemark{13} [erg cm s$^{-1}$]& $2.16_{-0.64}^{+1.75}$ & $3.69_{-0.15}^{+0.18}$ & $3.03_{-0.19}^{+0.89}$ & $1.64_{-0.16}^{+2.03}$ & $4.21_{-0.38}^{+0.22}$ & $3.79_{-0.68}^{+0.58}$ & $1.87_{-0.74}^{+1.32}$\\
\texttt{relxillCp} & $N_{\rm rel}$\tablenotemark{14} $(10^{-3})$ & $3.26_{-0.01}^{+0.01}$ & $6.02_{-0.01}^{+0.01}$ & $8.22_{-0.01}^{+0.02}$ & $7.00_{-0.02}^{+0.04}$ & $5.51_{-0.01}^{+0.02}$ & $0.7_{-0.4}^{+2.2}$ & $18.31_{-0.01}^{+0.01}$\\
\texttt{xillverCp} & $N_{\rm xil}$\tablenotemark{15} $(10^{-3})$ & $2.05_{-0.01}^{+0.01}$ & $1.84_{-0.01}^{+0.01}$ & $0.05_{-}^{+}$ & $2.50_{-0.01}^{+0.02}$ & $2.16_{-0.01}^{+0.01}$ &  $<1.44$ & $<0.01$\\
\midrule
$2-10$ keV Flux\tablenotemark{16} & [10$^{-8}$erg cm$^{-2}$ s$^{-1}$] & 0.31 & 0.50 & 0.62 & 0.63 & 0.64 & 0.76 & 1.07\\
Luminosity & $L/L_\mathrm{Edd}~(\%)$\tablenotemark{17} & 7.1 & 10.2 & 12.01 & 13.6 & 14.4 & 11.9 & 15.6\\
Reflection strength & $R_\mathrm{s}$\tablenotemark{18} & $0.07_{-0.03}^{+0.08}$ & $0.13_{-0.03}^{+0.02}$ & $0.34_{-0.08}^{+0.16}$ & $0.21_{-0.10}^{+0.09}$ & $0.17_{-0.06}^{+0.08}$ & $0.07_{-0.03}^{+0.05}$ & $0.32_{-0.11}^{+0.09}$\\
\midrule
 & {$\chi^{2}_{\nu}$}\tablenotemark{19} & $\frac{68.76}{63}=1.09$ & $\frac{64.97}{63}=1.03$ & $\frac{56.41}{63}=0.90$ & $\frac{50.78}{62}=0.82$ & $\frac{49.94}{62}=0.81$ & $\frac{136.26}{120}=1.14$ & $\frac{62.37}{62}=1.01$
\\ 
[1ex]
\enddata
\tablecomments{$^{1}$ Hydrogen column density; $^{2}$ Dimensionless spin parameter of the black hole ($a_{\star}$ = $cJ/GM^2$, where $J$ is the angular momentum of the black hole); $^{3}$ Disk inclination; $^{4}$ Iron abundance of the material in the accretion disk; $^{5}$ Reflection fraction, defined as the ratio of the reflected flux to the flux in the power-law component, in 20--40 keV band; $^{6,13}$ $\log$ of the ionization parameter ($\xi$) of the accretion disk, where $\xi=L/nR^{2}$, with \textit{L} as the ionizing luminosity, \textit{n} as the gas density and \textit{R} as the distance to the ionizing source; $^{7}$ Temperature at the inner disk; $^{9}$ Asymptotic power-law photon index; $^{8}$ Normalization of {\tt diskbb}, given by $N_{\rm disk}=(R_{\rm in}/D_{10})^2\cos{\theta}$, where $R_{\rm in}$ is the apparent inner disk radius (in km), $D_{10}$ is the distance to the source (in units of 10 kpc), and $\theta$ is the angle of inclination of the disk (in deg) \citep{1998PASJ...50..667K}; $^{10}$ Electron temperature determining the high energy roll-over; $^{11}$ When the normalization of {\tt nthComp} is unity, the corresponding model flux is 1 photon keV$^{-1}$cm$^{-2}$s$^{-1}$ at 1 keV; $^{12}$ Inner radius of the accretion disk; $^{14,15}$ Normalization of {\tt xillverCp} ($N_{\rm xil}$) is defined such that for an incident spectrum with flux $F_X(E)$ incident on a disk with density $n_e$, and with ionization parameter $\xi$, the integral $\int_{0.1 keV}^{1MeV} F_X(E)dE$ is evaluated to be $10^{20}\frac{n_e\xi}{4\pi}$. For {\tt relxillCp}, the definition of the normalization ($N_{\rm rel}$) is identical to that of {\tt xillverCp}, except that the flux reaching the observer is modified due to relativistic effects \citep[see][]{Dauserb}; $^{16}$ Unabsorbed flux (For reference, 1 Crab = $2.2\times10^8$ erg cm$^{2}$s$^{-1}$ in 2--10 keV band); $^{17}$ Luminosity of the source in terms of its Eddington limit, computed assuming the mass of the black hole to be $M = 10M_{\odot}$, a distance of $D = 8$ kpc (corresponding to $L_\mathrm{Edd}=1.25\times10^{38}$ erg s$^{-1}$), and from the 0.01--100 keV fluxes derived from the best-fit parameters; $^{18}$ Reflection strength, defined to be the ratio of the observed fluxes of the reflected component ({\tt relxillCp}) and the incident component ({\tt nthComp}) in the 20--40 keV energy band; $^{19}$ Reduced $\chi^2$, defined as the ratio of the best fit $\chi^2$ to the number of degrees of freedom ($\nu$).}
\label{tab:2004--2005}
\end{deluxetable*}

The 2004--2005 outburst (golden-brown color in Figure \ref{fig:HID}) is of particular interest because, it exhibits a different hysteresis loop in the HID, compared to the other outbursts. In this outburst, the hard to intermediate state transition occurs at a luminosity which is 2-3 times lower than what is observed in the case of the other outbursts (see the luminosity row in Tables \ref{tab:2002--2003} and \ref{tab:2004--2005}). The physical mechanisms that trigger the hysteresis observed in outbursts, and particularly, the anomalous behavior of the 2004--2005 outburst is not very well understood. This leaves us with the question of whether the global disk parameters including the inner disk radius differ in any way across the bright intermediate states of this outburst, from the previously analyzed one (2002--2003). 

For our exploratory analysis, we choose seven observations across the bright intermediate states of this outburst (see Table \ref{tab:observation_log}). The overall model set-up and the analysis procedure is identical to what is described in \S \ref{subsubsec:2002--2003} for the 2002--2003 outburst. The best fit parameters and the 90\% confidence intervals obtained by fitting the 2004--2005 observations with our reflection model are listed in Table \ref{tab:2004--2005}. The spectral data, total fitted model, the individual model components and the fit residuals are plotted in Figure \ref{fig:spectra}b.

Very much like the 2002--2003 outburst, the harder spectra of the 2004--2005 outburst ($HR\simeq 0.8,0.72,0.65$) do not exhibit the need for a blackbody component. Therefore, we freeze the normalization parameter of the \texttt{diskbb} component to 0, and fit for rest of the parameters. This restricts us from constraining the temperature of the inner disk for the hardest three observations. However, during the intermediate to soft states, the disk temperature of the 2004--2005 outburst is also consistently lower than the 2002--2003 outburst, which makes sense if the accretion rate ($\dot{M}$) is higher in the 2002--2003 outburst, although $\dot{M}$ need not be the only factor. We find that except for the coronal electron temperature ($kT_{\rm e}$), all the other parameters of the disk trace a similar trend across the bright intermediate states of the 2002--2003 outburst. Throughout this transition, $kT_{\rm e}$ values are seen to be unconstrained, with its lower limits found to be $>100-200$ keV during the hard to intermediate states ($HR\simeq 0.8-0.48$), and to be $>10$ keV during the softer states ($HR\simeq 0.3-0.05$).

\begin{figure*}[ht] \centering
\includegraphics[width=1\linewidth]{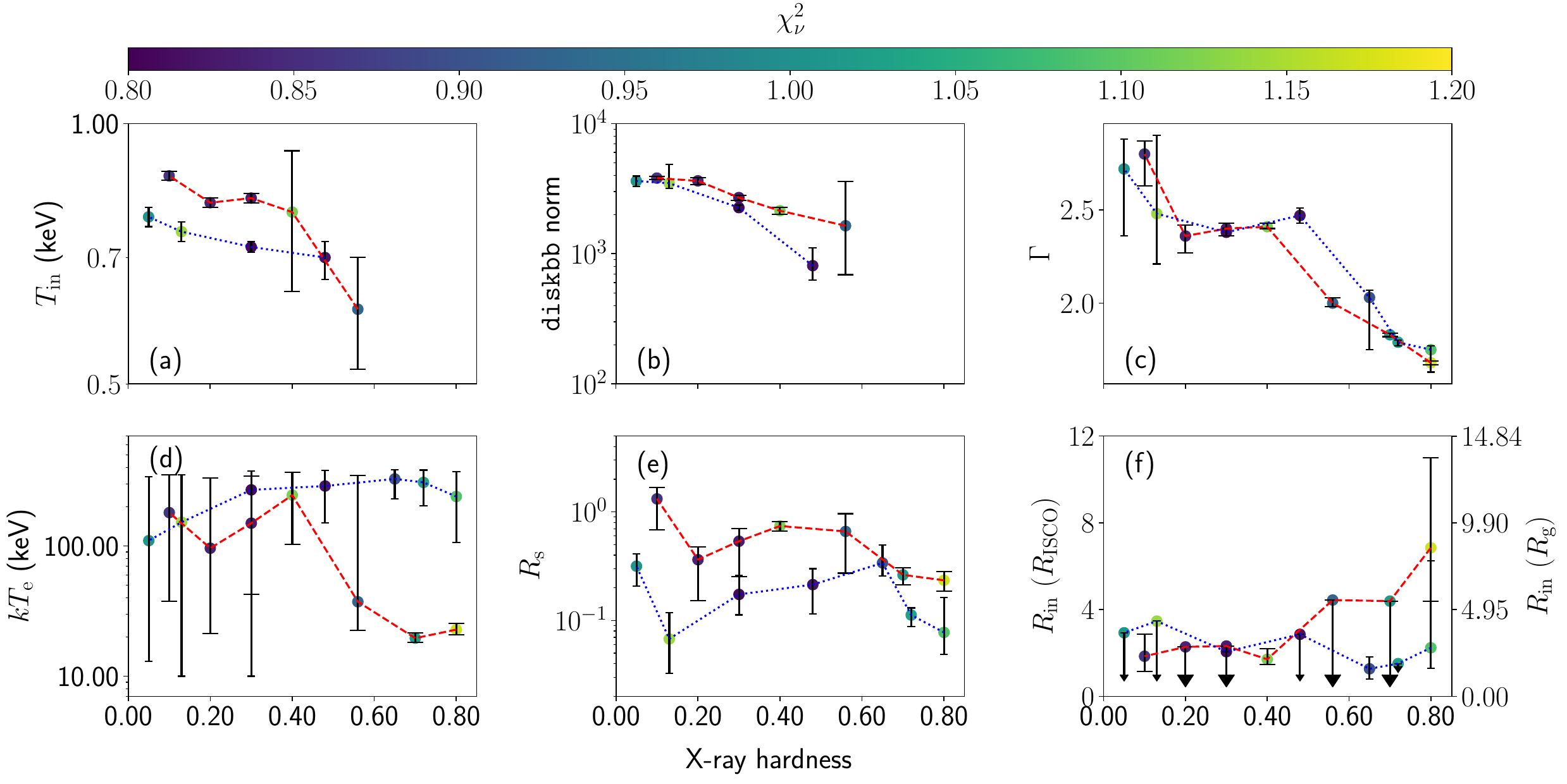}
\caption{Evolution of some of the free parameters of the reflection model: \texttt{TBabs(diskbb+nthComp+relxillCp+xillverCp)} across the bright intermediate states of GX 339--4's considered outbursts. The markers correspond to the median of the values obtained from MCMC probability distribution, and the error bars correspond to 90\% confidence interval. The best fit data points corresponding to the 2002--2003 outburst are interlinked by red dashes and the 2004--2005 outburst by blue dots. Plotted on the horizontal axis is the X-ray hardness, defined as the ratio of the source counts at $8.6-18$ keV to the counts at $5.0-8.6$ keV energy band. The vertical axes represent (a) inner disk temperature ($T_\mathrm{in}$); (b) \texttt{diskbb} normalization parameter ($N_{\rm disk}$); (c) power-law photon index ($\Gamma$); (d) electron temperature ($kT_{\rm e}$); (e) Reflection strength ($R_{\rm s}$), as defined in \S\ref{subsubsec:2002--2003}; (f) inner disk radius (in units of $R_{\rm ISCO}$ (left ordinate) and gravitational radius, $R_{\rm g}$ (right ordinate) for a black hole spin of $a_{\star}=0.998$). Each marker represents an individual observation, color coded by the reduced $\chi^2$ from the fit to the model. The arrows at the end of the error bars indicate that the particular model parameter is unconstrained in that direction.}
\label{fig:reflection_evolution}
\end{figure*}

Throughout the bright intermediate states of the 2004--2005 outburst, we find that the inner disk radius never recedes away by more than 6.23 $R_{\rm ISCO}$ (7.7 $R_{\rm g}$ at $HR\sim0.8$), and the evolutionary trend at later times ($HR<0.8$) is seen to be very much akin to the 2002--2003 outburst (see Figure \ref{fig:reflection_evolution}). 

For this outburst, the reflection strength is seen to be the weakest ($R_{\rm s}=0.07_{-0.03}^{+0.08}$) during the bright hard state ($HR\simeq0.8$), and during the bright intermediate states ($0.7\lesssim HR\lesssim0.2$), $R_{\rm s}$ is seen to oscillate between 0.10 and 0.12. This is followed by a steep rise in its value, culminating at $R_{\rm s}=0.32_{-0.11}^{+0.09}$, by the time the system reaches the bright soft state ($HR\sim 0.1$). This is a $2.1\sim 8.1$ fold rise in the reflection strength, from its value during the inception of bright intermediate state transition. This analysis demonstrates that although the reflection strength follows a similar evolutionary pattern across the bright intermediate states of the 2002--2003 and 2004--2005 outbursts, its magnitude is higher for the outburst with higher state-transition luminosity (2002--2003) and lower for the outburst with lower state-transition luminosity (2004--2005).

See Figure \ref{fig:reflection_evolution} for a visual comparison of the evolution of key parameters from the reflection spectrum analysis across the bright intermediate states of the 2002--2003 and 2004--2005 outbursts. 

\subsubsection{Markov Chain Monte Carlo (MCMC) analysis}\label{subsec:MCMC}

In order to substantiate the parameter values derived from the complex reflection model, we perform a full-fledged Markov Chain Monte Carlo (MCMC) statistical analysis for all the 14 reflection fits (see \S 4 of \citealt{2012ApJ...745..136S} for more details). The parameter space is explored in the MCMC analysis for each spectral fit with 50 walkers (distinct chains), which explore the parameter space in a sequence of 900,000 affine-invariant stretch-move steps \citep{2013PASP..125..306F}. After having been initialized in a cluster distributed around the best fit parameter values, the first 30\% of the elements traversed by each walker are discarded as the `burn-in' phase, during which the chain reaches its stationary state. Autocorrelation length, which is the interval over which the chain forgets its previous location, for each walker, is computed with respect to the lag in the length traversed. A typical value of the autocorrelation length is found to be $\sim$10,000 elements ($\sim$ 10\%), which corresponds to a net number of independent samples of the parameter space to be 90 per walker. All those walkers, whose autocorrelation values do not reach zero before it traverses $\sim$10\% of the total chain length are considered to be not converged, and are discarded. From the full distribution, we obtain a probability distribution for any given set of parameters of interest by marginalizing over all the other variables that are outside that set. Logarithmically-flat priors are adopted for all model components' normalization parameters. To regularize the space over which the parameters are sampled from a finite interval to a real line, each parameter is remapped using a logit transformation.

In Appendix \ref{appendix:mcmc}, we show the contour maps and probability distributions for the set of most relevant physical parameters, derived using the MCMC analysis. For the cases of $HR\simeq0.80$ and 0.70 in the 2002--2003 outburst, and $HR\simeq 0.8, 0.72$ and 0.65 in the 2004--2005 outburst, the {\tt diskbb} norm parameter is not shown in the contour maps because, the spectra are hard enough to require any disk blackbody component to describe it, and hence the parameter normalization is frozen to zero there (see \S \ref{subsubsec:2002--2003} and \S \ref{subsubsec:2004--2005}). For all the observations modeled with a free {\tt diskbb} normalization parameter, the parameter space spanned by the walker clearly indicates a strong anti-correlation between $T_\mathrm{in}$ and the component's norm, while $\Gamma$ and $R_{\rm in}$ show little dependence on other parameters (see Figures \ref{fig:corner_2002--2003} and \ref{fig:corner_2004--2005}). However, the hard state observations do reveal a degeneracy in the value of the spectral index. The MCMC analysis performed here also helps us in revealing the bimodal distribution of parameter values, seen in the contour plots as two different cluster of points ($HR\simeq0.8, 0.65, 0.48$ in Figure \ref{fig:corner_2004--2005}). Despite the fact that some parameters exhibit bimodal distributions, we note that they do not necessarily provide the best physical interpretation of the data, and are just indicative of a few walkers being converged at local minima. The median value of each free parameters' distribution, and the 90\% confidence interval derived from the chains are listed in Tables \ref{tab:2002--2003} \& \ref{tab:2004--2005}.

\subsection{Fitting without the \textsc{relxill} model}\label{subsec:comprehensive}
Our analysis of the selected spectra from the bright intermediate states using the standard (blurred+unblurred) reflection model has indicated a relatively stable state of the inner edge of the accretion disk. Yet, continuum parameters like $T_{\rm in}$, \texttt{diskbb} norm and $kT_{\rm e}$ are seen to be changing, paralleled with the softening of the spectra (see Tables \ref{tab:2002--2003} \& \ref{tab:2004--2005}). In this section, we perform an alternate analysis, which, although isn't rigorous on the reflection component, can probe the continuum emission from the disk and corona to greater profundity. This analysis employs a self-consistent Comptonized accretion disk model that also accounts for the scattering of the disk photons by the corona. Therefore, we note that the Comptonization continuum provided by this model is not identical to that of the previously employed reflection model, and we value this distinction. With this setup, our full model of the Comptonized accretion disk in this case is a convolution of the multi-temperature disk black body (\texttt{diskbb}) component with an empirical \texttt{simplcut}\footnote{\url{http://jfsteiner.com/simplcut/}} model \citep{2017ApJ...836..119S}, which convolves a Comptonization Green's function with a seed photon spectrum, thereby redirecting the photon from the seed distribution (described by \texttt{diskbb}) into a power-law. In this model, we use a simple \texttt{Gaussian} component to account for the Fe K emission, and the \texttt{smedge} component for the smeared absorption edge. The \texttt{TBabs} and the \texttt{smedge} components are handled the same way as mentioned in \S \ref{subsec:ratio}, and we fit for the rest of the free parameters of the model. Therefore, the final model setup in \textsc{xspec} notation is: \texttt{TBabs*smedge(simplcut$\otimes$diskbb+Gaussian)}. We perform this analysis to retrieve meaningful coronal electron temperatures and hard-state $T_{\rm in}$, which the previous reflection analysis could not yield a constraint on, and to independently estimate the inner disk radius from the {\tt diskbb} norm.

\begin{figure*}[ht] \centering
\includegraphics[width=1\linewidth]{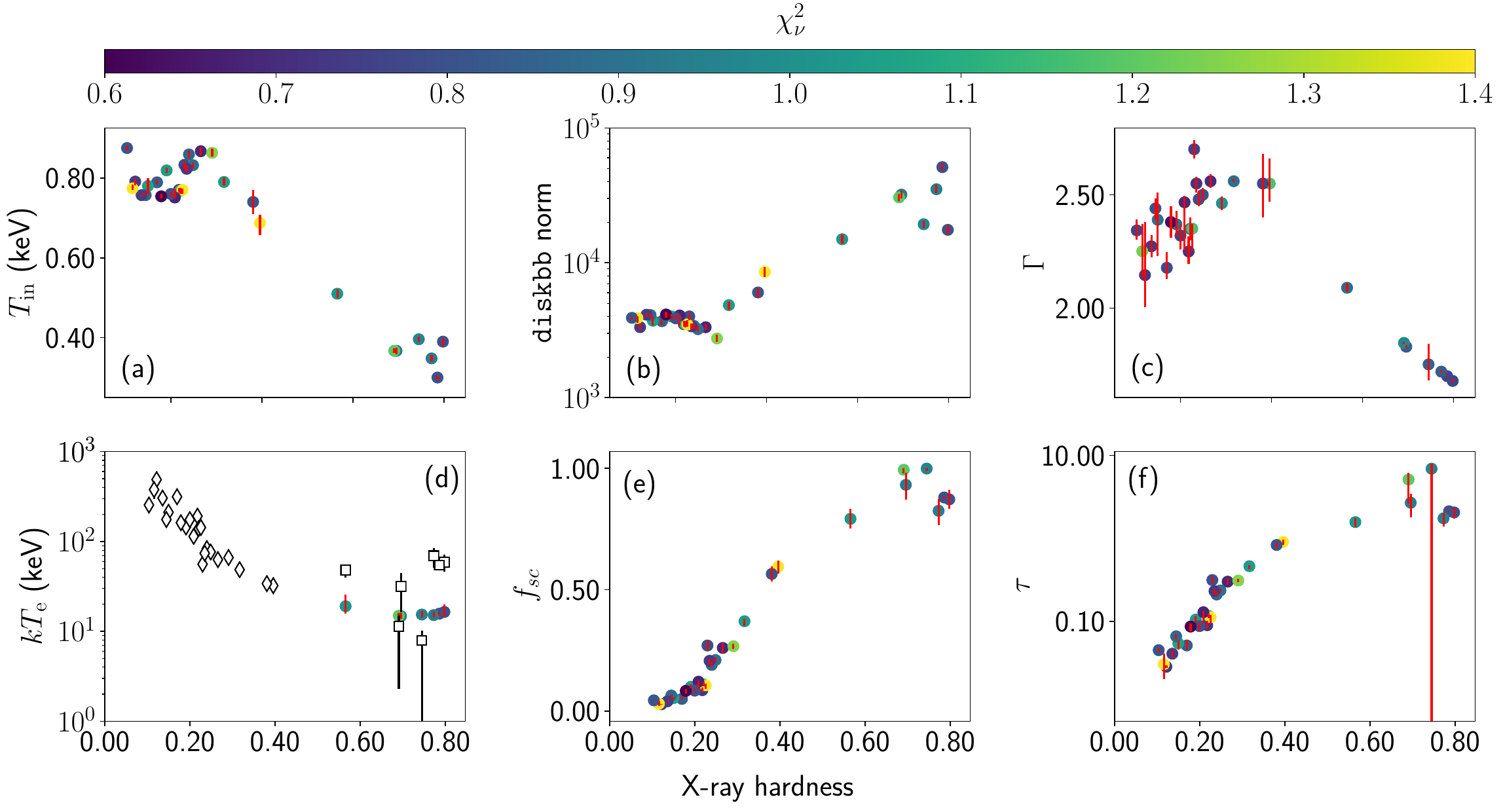}
\caption{Evolution of the free parameters of the continuum model: \texttt{TBabs*smedge(simplcut$\otimes$diskbb+Gaussian)} across the bright intermediate states of the 2002--2003 outburst of GX 339--4. Plotted on the horizontal axis is the X-ray hardness, defined as the ratio of the source counts at $8.6-18$ keV to the counts at $5.0-8.6$ keV energy band. The vertical axes represent (a) inner disk temperature ($T_\mathrm{in}$); (b) \texttt{diskbb} norm; (c) power-law photon index ($\Gamma$); (d) electron temperature---colored circles denote the values obtained from best-fit to the data (wherever constrained values are available), and open square and rhombus markers denote the analytically calculated values for high and low optical depth regime, respectively (see \S\ref{subsubsec:kTe}); (e) scattered fraction ($f_{sc}$); (f) optical depth ($\tau$) for a covering fraction of unity. Each marker represents an individual observation, color coded by the reduced $\chi^2$ from the fit to the model}
\label{fig:simplcut}
\end{figure*}

With this model, we fit a series of observations monotonically decreasing in their hardness from the bright hard to soft states of the 2002--2003 outburst. The relatively simple nature of this model allows us to include more observations from the bright soft state than the reflection analysis. The evolution of the key parameters of the disk and the corona viz., $T_{\rm in}$, \texttt{diskbb} norm, photon power-law index ($\Gamma$), the fraction of seed photons scattered into power-law distribution ($f_{sc}$), optical depth ($\tau$), and the coronal electron temperature ($kT_{\rm e}$) are depicted in Figure \ref{fig:simplcut}. The optical depth is derived by assuming a covering fraction of unity using the relation $\tau=-\ln{(1-f_{sc})}$ \citep{2017ApJ...836..119S}, and depicted along with the other parameters in Figure \ref{fig:simplcut}f, for the sake of completion.

The increasing nature of $T_{\rm in}$ with softening of the spectra is seen to persist until the spectral hardness reaches a value of $HR\simeq 0.3$, followed by a saturation of its value at $\sim 0.75$ keV until the bright soft state. On the other hand, the norm of the \texttt{diskbb} parameter, being degenerate with $T_{\rm in}$, traces a decreasing trend as the spectra softens until the spectral hardness reaches a value of $HR\simeq 0.3$. This trend is followed by a saturation of its value at $\sim 3\times 10^3$ until the bright soft state. The power-law index, which is flattest during the hard state ($\Gamma \sim1.6$), reaches a value of $\Gamma \sim 2.4$ during the bright soft state, passing by the spectral hardness value of $HR\simeq 0.3$, where it reaches its maximum value of $\Gamma \sim 2.7$. This value of spectral hardness corresponds to what is known as the `steep power-law state' as defined by \cite{2006csxs.book..157M}, and first observed in this source by \cite{1991ApJ...383..784M}. Unlike the previously mentioned parameters, we do not observe any stark kink in the evolution of the scattered fraction ($f_{sc}$) and therefore $\tau$, at the steep power-law state. They exhibit a strong positive correlation with the spectral hardness throughout the bright intermediate states. The coronal electron temperature ($kT_{\rm e}$) can be constrained with this model only during the bright hard to intermediate states $0.8<HR<0.56$ (see Figure \ref{fig:simplcut}). The value of $kT_{\rm e}$ obtained from this analysis ($15\sim 20$ keV in $0.8<HR<0.56$, unconstrained elsewhere) closely resembles the value of $kT_{\rm e}$ obtained from the reflection analysis of the 2002--2003 outburst (see Table \ref{tab:2002--2003}). Analysis of the reflection spectrum convolved with the self-consistently Comptonizing \texttt{simplcut} model is deferred for a future work.

\section{Discussion} \label{sec:discussion}

\subsection{Inner disk radius ($R_{\rm in}$)} \label{subsec:Rin}

A number of studies \citep[e.g.,][]{2010MNRAS.407.2287D, 2016MNRAS.458.2199B} have reported a broadening in the relativistic Fe K line as the spectra becomes softer, and therefore, have suggested that the inner radius correlates with the X-ray $HR$. Should this be the case, the signatures of increase in line width with softening of the spectra should also be seen with \textit{RXTE}/PCA\footnote{See Appendix \ref{appendix:rxte} for a test of the potential of \textit{RXTE}/PCA to detect narrow Fe K emission line, corresponding to large disk truncation.}. The data/model ratio plot in Figure \ref{fig:Fe_ratio} shows the Fe K emission feature, plotted for three different representative values of $HR$ across the bright-intermediate states. Clearly, no significant broadening of the emission line can be seen, as the spectra softens, accompanied with the decrease in $HR$ value from 0.8 to 0.1. Should the inner disk's truncation be greatly decreasing as the system approaches soft state from bright hard state, we would observe, as a consequence, a steady broadening of the emission line due to gravitational red shift \citep{2000MNRAS.315..223R}. Our result therefore qualitatively indicates that the inner radius of the accretion disk, which is presumed to be the source of the Fe K emission line, is independent of the spectral hardness during the bright-intermediate states' transition. 

We then set out to quantitatively answer the question of where the inner disk truncates for the two outbursts we choose with different bright-intermediate states' transition luminosities (2002--2003 and 2004--2005 outbursts). In order to answer this, we perform an analysis of the reflection spectrum using the \textsc{relxill} family of relativistic reflection models. Note that the {\tt diskbb} norm ($N_{\rm disk}$) from the reflection fit is not indicative of the size of the disk, since it only models the unscattered (transmitted) disk emission. From the values of the $R_\mathrm{in}$ constrained from the analysis of the 2002--2003 and 2004--2005 outbursts (see Figure \ref{fig:reflection_evolution}) using the \texttt{relxillCp} component, we find that the inner truncation of the accretion disk upon reaching very close to ISCO by the onset of bright hard state \citep{2015ApJ...813...84G, 2018ApJ...855...61W, 2019arXiv190800965G}, remains at a steady value ($R_{\rm in}\sim R_{\rm ISCO}$) regardless of the luminosity at which the black hole undergoes its transition across the bright hard to soft state. A plausible mechanism that can explain the sustained disk truncation close to the ISCO overcoming the putative magnetically arrested state of the disk \citep{2003PASJ...55L..69N} is, by annihilation of the accumulated central fields by quasi-periodic inversion of poloidal field lines \citep{2009ApJ...702L..72I}. According to \cite{2009A&A...508..329M}, the hard to soft state transition luminosity of an outburst occurs at a relatively lower luminosity (as in 2004--2005 outburst), if the inner disk had remained stationary from the previous outburst (i.e., 2002--2003), plausibly due to re-condensation of matter from the ADAF \citep{1998tbha.conf..148N}. As per this picture, even though the inner edge of the outer disk could be highly truncated during quiescence and low/hard states, a weak inner disk might still be present during the low luminosity states, giving rise to small truncation radius as seen from our reflection study.

\subsection{Iron abundance, disk density and jets}

A recent study of the high-density reflection models by \cite{2019MNRAS.484.1972J} has indicated that the bright soft state spectra of GX 339--4 can be described by a near Solar iron abundance ($A_{\rm Fe}$). We nevertheless use the aforementioned super-Solar value, as the density of the disk at bright soft state is two orders of magnitude smaller than the low hard state. We note that the model flavor \texttt{relxillD} allows for a variable density of the disk. We however do not use it, as the current version of the model only allows for a fixed cut-off energy of 300 keV. That does not describe the varying properties of the spectra seen across the bright-intermediate states well, and appropriate model with variable cut-off energy is not readily accessible.

There have been models on the formation and evolution of a synchrotron-emitting, large-scale transient jet during the bright-intermediate state transition of GX 339--4 \citep{1991ApJ...374..741M}. Particularly,  the 2002--2003 outburst of this source began exhibiting large-scale jet during 2002 May 14 \citep{2004MNRAS.347L..52G}. In our selected observations (Table \ref{tab:observation_log}), this is also the duration when the source makes a transition from $HR=0.4\rightarrow0.3$ (steep power-law state). This transition, accompanied by a jet, is seen to give rise to several interesting signatures in the evolutionary trend of the source parameters. Such features include the inflections seen in the evolution of power-law photon index (Figures \ref{fig:reflection_evolution}c \& \ref{fig:simplcut}c), reflection strength (Figure \ref{fig:reflection_evolution}f), and the onset of saturation in the value of $T_{\rm in}$ and {\tt diskbb} norm (Figures \ref{fig:simplcut}a \&  \ref{fig:simplcut}b). Although it is perhaps easy to attribute the sudden changes seen in the source parameters to the emission of ballistic jets, the underlying physics of jet launching and its coupling with the disk/corona is still not well understood. However, from our results and previous discussion on the disk truncation (\S\ref{subsec:Rin}), we can at least argue that the advancement of the accretion disk towards the black hole need not be necessary for triggering the transient relativistic jet seen during the intermediate state transition in a microquasar like GX 339--4.

\subsection{Reflection strength ($R_{\rm s}$)} \label{subsec:Rs}

The approach we follow in order to parametrize the strength of the reflection is by calculating the ratio of the observed fluxes of the reflected component ({\tt relxillCp}) to that of the incident component ({\tt nthComp}) in the 20--40 keV energy band, a definition that has been used by \cite{2015ApJ...806..149K} and \cite{2015ApJ...811...51T}. This energy range is particularly of interest because, here the
reflection spectrum is independent of the ionization parameter $\xi$ (with no Fe K fluorescent emission feature), and it also encompasses the peak of the Compton hump, where the reflection spectrum is dominated due to Compton scattering. Another parameter quantifying reflection off the disk is the reflection fraction ($R_{\rm f}$), which is defined as the ratio of the coronal intensity that illuminates the disk to the coronal intensity that reaches the observer. A correspondence between $R_{\rm s}$ and $R_{\rm f}$ for different disk inclination and black hole spin values is given in \cite{2016A&A...590A..76D}.

Our results indicate a weak reflection during the bright hard state of the 2002--2003 and 2004--2005 outbursts. A quick reasoning would be to ascertain that the disk is truncated far from the black hole, thus resulting in a diminished reflection from the disk. However, the estimates of the inner disk truncation from our reflection analysis (see Tables \ref{tab:2002--2003} \& \ref{tab:2004--2005}) and other works \citep{2015ApJ...813...84G, 2018ApJ...855...61W, 2019arXiv190800965G} indicate otherwise, and we would like to emphasize here that such a weak reflection does not necessarily imply a disk which is disrupted, or truncated far from the black hole. Studies by \cite{1999ApJ...510L.123B} and \cite{2001MNRAS.326..417M} have shown that a non-static corona during the hard state of a black hole can lead to an attenuated reflection strength due to relativistic aberration. Another reason for this attenuation in $R_{\rm s}$ can be attributed to the large optical depth of corona seen in the hard state (see Figure \ref{fig:simplcut}f), which dilutes the amplitude of the Fe line due to the resultant increased Compton scattering \citep{2001MNRAS.328..501P, 2016ApJ...829L..22S}. This attenuation in $R_{\rm s}$ during the bright hard state is followed by an increasing trend in the reflection strength with the softening of the spectra, which is consistent with the results of \cite{2016ApJ...829L..22S}. This can be attributed to the `compactification' of the corona. In this paradigm, the ratio of the scale-height of the corona ($h$) to $R_{\rm in}$ decreases with decreasing spectral hardness \citep{2019Natur.565..198K}. Another possible scenario is when the radius of the co-rotating slab corona decreases, thereby leading to a softer emission \citep{2019ApJ...875..148Z}.

\subsection{Temperatures} \label{subsec:temperatures}
\subsubsection{Coronal electron temperature ($kT_{\rm e}$)}  \label{subsubsec:kTe}

The electron temperatures ($kT_{\rm e}$) in the corona and its confidence intervals cannot be constrained well from our analysis, especially during the softer states of both the 2002--2003 and 2004--2005 outbursts. However during the bright hard state, the $kT_{\rm e}$ values of the 2004--2005 outburst, having their lower limits at $\sim 100$ keV are higher than the upper limit of $kT_{\rm e}$ values at the corresponding hardness value of the 2002--2003 outburst. This can be attributed to the inverse Compton cooling (IC) effect. The IC cooling power is given by, 
\begin{equation}
P_{\rm IC}=\frac{4}{3}\sigma_{T}c\beta^2\gamma^2U_{rad},
\end{equation}
where $\sigma_T$ is the Thomson cross section, $\beta$ is the velocity of the electrons in units of speed of light (c), $\gamma$ is the Lorentz factor, and $U_{rad}$ is the energy density of the incident radiation. The bright-intermediate states of the 2002--2003 outburst are on an average $\sim 3$ times more luminous than the 2004--2005 outburst, leading to a higher value of $U_{rad}$. This results in a more effective cooling of the corona in the 2002--2003 outburst compared to the 2004--2005 outburst. A reversal in the trend of $kT_{\rm e}$ is observed between the two outbursts, with the softening of spectra, as seen from Figure \ref{fig:reflection_evolution}d. However, we do not find this trend credible enough, owing to the unconstrained nature of the $kT_{\rm e}$ values during the softer states. We note here that the overall trend of $kT_{\rm e}$ of the 2002--2003 outburst from reflection fitting is similar to that obtained from the non-reflection model (see Figures \ref{fig:reflection_evolution}d and \ref{fig:simplcut}d).

In an isothermal corona, the pair/electron-ion plasma is known to Compton up-scatter the thermal disk photons, thereby producing the observed power-law continuum. For known values of the power-law index ($\Gamma$) and the electron temperature in the plasma ($kT_{\rm e}$), the Thomson optical depth ($\tau$) of the corona at different regimes can be estimated using the empirical relations \citep{1983ASPRv...2..189P, 1987ApJ...319..643L, 1993ApJ...413..507H},

\begin{subnumcases}{\Gamma=}
   -\frac{1}{2}+\sqrt{\frac{9}{4}+\frac{1}{\theta_e\tau(1+\tau/3)}} & $:\tau \gtrsim 1$ \label{eq:Gamma1}
   \\
   1 + \frac{-\ln{\tau} + 2/(3+\theta_e)}{\ln{12\theta_e^2 + 25\theta}} & $:\tau \lesssim 1$  \label{eq:Gamma2}
\end{subnumcases}

where $\theta_e$ is dimensionless temperature parameter given by $\theta_e=kT_{\rm e}/m_ec^2$, where $m_ec^2=511$ keV is the rest mass energy of electron. The unconstrained nature of $kT_{\rm e}$ does not allow for the estimation of the optical depth, $\tau$ from eqs. \ref{eq:Gamma1} and \ref{eq:Gamma2}. However, the \texttt{simplcut} component of our model (see \S \ref{subsec:comprehensive}) provides us with the scattered fraction ($f_{sc}$), which is related to the optical depth $\tau$ by, $\tau=-\ln{(1-f_{sc})}$, assuming a uniform density corona with a covering fraction of unity. The covering fraction of corona for a realistic accretion disk is expected to be $<1$ \citep{2015MNRAS.448..703W}, and therefore our assumption places a lower limit on $\tau$. Using the analytical estimates of $\tau$, and the best fit values of $\Gamma$, we retrieve back the electron temperatures using eqs. \ref{eq:Gamma1} and \ref{eq:Gamma2}, for different values of hardness across the bright-intermediate states of the 2002--2003 outburst (see Figure \ref{fig:simplcut}d). The largest fraction of soft seed photons are produced in the softer state (see Figures \ref{fig:reflection_evolution}a \& \ref{fig:simplcut}a). Despite a large value of $U_{rad}$ during the softer states, the analytical estimate of $kT_{\rm e}$ is seen to grow higher, as the system gets softer. This is not unexpected, as the coronal electron temperature increases with decrease in optical depth (lesser scattering), which is limited by the creation of of $e^{\pm}$ pairs from magnetic reconnection events \citep{1983MNRAS.205..593G, 1993ApJ...413..507H, 2017ApJ...850..141B}.

We note here that all the coronal properties we place a constraint on, are based on the assumption of seed photons being uniformly distributed and isotropic. For anisotropic cases, we refer readers to the recent results of \cite{2019ApJ...875..148Z}.

\subsubsection{Inner disk temperature ($T_{\rm in}$) and \\the spectral hardening factor ($f_{col}$)}  \label{subsubsec:Tin}

The fact that GX 339--4 spends a couple of months in the bright-intermediate states without exhibiting large changes in luminosity, but with just spectral hardness, also suggests a relatively stable configuration of the disk, which is plausible if the disk is near or at the ISCO. If this is the case, then the impending transition to the bright soft state should be triggered by a rather different physical process than any substantial radial motion of the disk. One plausible candidate is the surge in the accretion rate \citep{2000MNRAS.313..193M, 2008NewAR..51..733N, 2013ApJ...769..156S}. For a standard optically thick accretion disk, the blackbody temperature of the disk ($T_{bb}$) at a given radius, $r$ is related to the accretion rate $\dot{M}$ by the relation,

\begin{equation} \label{eq:accretion_rate}
\sigma T_{bb}^4 = \frac{GM}{r^3}\frac{3\dot{M}}{8\pi}\bigg[1-\bigg(\frac{r_{in}}{r}\bigg)^{1/2}\bigg],
\end{equation}
where $M$ and $r_{in}$ denote the mass of the accreting body and the inner radius of the accretion disk. Eq. \ref{eq:accretion_rate} tells that for a relatively stable $r_{in}$, an increase in the accretion rate (as the spectra softens) should result in an increase in the inner disk temperature \citep[possibly amplified by irradiation coupling;][]{1998MNRAS.293L..42K}. We note here that our results manifest the same (see Figure \ref{fig:simplcut}). In fact, the increasing trend of $T_{\rm in}$ from our analysis with decreasing spectral hardness, display an agreement between the models with and without the reflection components (see Figures \ref{fig:reflection_evolution} and \ref{fig:simplcut}).

The inner disk temperature ($T_{\rm in}$) returned by the \texttt{diskbb} component is the color temperature, which is related to the effective temperature ($T_{\rm eff}$) of the disk by $T_{\rm in}=f_{col}T_{\rm eff}$, where $f_{col}$ is the color-correction (a.k.a spectral hardening) factor \citep{1995ApJ...445..780S}. \cite{2013MNRAS.431.3510S} had demonstrated that physically reasonable changes in the phenomenological color-correction factor ($f_{col}$) can also provide a plausible description for the hard to soft state transition, without needing to invoke radial motion of the inner accretion disk \citep{2019ApJ...874...23D}.  Assuming a canonical value of $f_{col}=1.7$ at the bright soft state, we estimate the temperature dependent values of $f_{col}$ for varying spectral hardness, using the scaling relation, $f_{col}\propto T_{\rm in}^{1/3}$ \citep{2006ApJ...647..525D}. The $T_{\rm in}$ values used in this calculation are extracted from the model: \texttt{TBabs*smedge(simplcut$\otimes$diskbb+Gaussian)}. We use the values from this model, in order to arrive at $R_{\rm in}$ measurements independent of the results from the reflection model. Note that the {\tt diskbb} norm from the reflection fit is not indicative of the size of the disk, since it only models the unscattered (transmitted) disk emission. This is not the case with the {\tt diskbb} norm from this alternate analysis involving the {\tt simplcut} model. The normalization of the {\tt diskbb} component is given by, $N=(r_{\rm in}^{\rm app}/D_{10})^2 \cos{\theta}$, where $r_{\rm in}^{\rm app}$ is the `apparent inner disk radius', $D_{10}$ is the distance to the black hole in units of 10 kpc, and $\theta$ is the inclination value of the disk. Using the relationship between the apparent and the true inner disk radius \citep{1998PASJ...50..667K}, the {\tt diskbb} normalization (N) can be rewritten in terms of $f_{col}$ as,
\begin{equation} \label{eq:norm}
N=\bigg(\frac{r_{in}}{\Xi f_{col}^2 D_{10}}\bigg)^2 \cos{\theta},
\end{equation}
where $r_{in}$ is the true inner disk radius (in units of km) and $\Xi$ is the relativistic color correction factor (of the order of unity). We would like to emphasize here that a negligible value of the {\tt diskbb} normalization, or a non-requirement of the disk quasi-blackbody component in the spectrum (as seen in the hard state spectra of our reflection analysis) does not necessarily imply that the accretion disk is highly truncated, or even absent. A likely interpretation for this is that the cool thermal photons emitted by the disk are almost entirely Comptonized in a $\tau \gtrsim 1$ corona (see Figure \ref{fig:simplcut}f), and so the transmitted (unscattered) portion of that emission is indiscernibly weak.  Note that this does not prevent one from being able to place one-sided constraint of {\tt diskbb} parameters with the {\tt simplcut} model. Moreover, \textit{RXTE}/PCA has its low-energy cut-off at $\sim$3 keV, allowing only the Wien tail of the blackbody spectrum in the observable band. The blackbody component during the hard state of GX 339--4, primarily dominant at energies below PCA's energy sensitivity has been observed with other instruments \citep{2006ApJ...653..525M, 2008ApJ...680..593T, 2008MNRAS.387.1489R, 2015A&A...573A.120P, 2016MNRAS.458.2199B}. 

\begin{figure}[t!]
\includegraphics[width=1\linewidth]{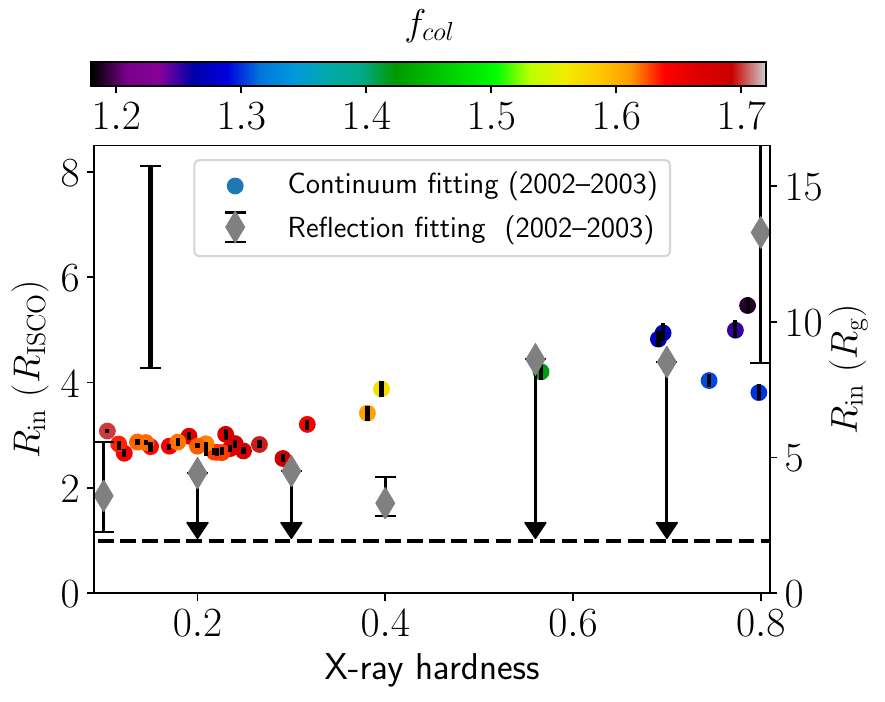}
\caption{Independent measurements of the evolution of the inner disk radius ($R_{\rm in}$) from the continuum fitting method (round markers), in comparison with the results from the reflection method (grey rhombus markers), plotted against the X-ray hardness. The inner disk radii are plotted in units of $R_{\rm ISCO}$ (left ordinate) and gravitational radius ($R_{\rm g}$; right ordinate), for a black hole spin of $a_{\star}=0.95$. The vertical error bars (on the marker) represent the propagated statistical uncertainties from the model fitting, and the systematic uncertainty arising from the ignorance of the system's intrinsic parameters is represented by the large vertical error bar with cap, at the top left region of the Figure. The round markers are color-coded with the value of the spectral hardening factor ($f_{col}$), which is used to calculate $R_{\rm in}$.}
\end{figure}\label{fig:Rin_fcol}

As a next step, we plug in previously estimated values of $f_{col}$ from the scaling relation into eq. \ref{eq:norm}, and retrieve the evolution of inner disk radius assuming the following about the system's intrinsic parameters: $M=8\pm2 M_{\odot}$, $D_{\rm 10} = 0.8\pm0.2$, inclination, $\theta$ = $45\pm15^\circ$ and the dimensionless spin parameter to be $a_{\star}=0.950\pm0.049$ (see \S\ref{sec:intro}). Figure \ref{fig:Rin_fcol} shows the evolution of the inner disk radius independently measured from the continuum fitting method, and the corresponding value of $f_{col}$ used at different hardness values. From Figure \ref{fig:Rin_fcol}, we see that the radius of the inner disk's truncation obtained via continuum fitting method stays at $\lesssim$4 $R_{\rm ISCO}$ across the bright intermediate states. These independent measurements of $R_{\rm in}$ are in excellent agreement with the inner disc truncation obtained from reflection analysis (compare circle and diamond markers in Figure \ref{fig:Rin_fcol}), further bolstering our results. 

We note here that changes seen in the norm of the {\tt diskbb} component (as in Figure \ref{fig:simplcut}b) do not necessarily imply a change in the inner disk radius, due to the existing degeneracy between $f_{col}$ and $R_{\rm in}$. On the other hand, previous works by \cite{2013ApJ...769...16R} and \cite{2013MNRAS.431.3510S} have demonstrated that order of magnitude changes seen in the \texttt{diskbb} normalization can be explained by merely invoking changes in the spectral hardening factor. Furthermore, with a steady increase in $f_{col}$ seen with the softening of the spectra, it is easier to conceive that the bright hard to soft state transition is perhaps triggered by the magnetic or surface properties of the accretion disk, rather than by a sudden appearance of the disk. Particularly, the strong toroidal magnetic fields generated by the magnetorotational instability \citep{velikhov1959stability, 1960PNAS...46..253C, 1991ApJ...376..214B} would cause the disk to thicken, consequently decreasing the viscous inflow time \citep{2007MNRAS.375.1070B, 2008A&A...490..501J}. The existence of strong correlation between $f_{col}$ and the viscosity parameter \citep[$\alpha$;][]{1973A&A....24..337S} in the inner regions of the disk is studied in \cite{1995ApJ...445..780S}. This dependence can be explained on the grounds of a gradual increase seen in the electron temperature as the disk becomes optically thin (see Figure \ref{fig:simplcut}f), with an increase in $\alpha$ due to changes in $\dot{M}$. We refer readers to the works of \cite{2005ApJ...621..372D, 2006ApJ...647..525D} and \cite{2019ApJ...874...23D} for a more comprehensive discussion on the dependence of $f_{col}$ on other physical processes. 

\section{Summary} \label{sec:summary}

In this paper, we perform an analysis of the reflection spectrum of the black hole X-ray binary GX 339--4, as observed by \textit{RXTE} during the transition from the bright hard to soft states of the 2002--2003 and 2004--2005 outbursts. We also perform an analysis of the continuum spectra from the bright intermediate state transition by employing a self-consistent Comptonized accretion disk model accounting for the scattering of the disk photons by the corona, to arrive at the disk and coronal properties independent of the reflection modeling, and compare one against each other. Our chief results are the following:

\begin{enumerate}
\item Strong features of broad Fe K emission are seen in the data across the bright intermediate state transition, and the profile of the Fe K is mostly insensitive to the Comptonization model used (\S\ref{subsec:ratio}).

\item Qualitatively, the width of the Fe K emission feature is seen to be nearly constant across the bright hard to soft state transition---indicating a quasi-static inner disk truncation radius (\S\ref{subsec:ratio}).

\item By fitting the data with the \textsc{relxill} family of relativistic reflection models, we conclude that the inner edge of the accretion disk reaches $\sim 14 R_{\rm g}$ by the onset of bright hard state, and the truncation remains at $\lesssim 5.5 R_{\rm g}$ across the bright-intermediate state transition (\S\ref{subsubsec:2002--2003}).

\item Regardless of the luminosity at which the source undergoes the bright hard to soft state transition in this source (as in the lower luminous 2004--2005 outburst), the inner disk truncation is seen to be behaving similarly, as mentioned above (\S\ref{subsubsec:2004--2005}).

\item While disk and coronal parameters like disk black body normalization, power-law photon index ($\Gamma$), $R_{\rm in}$, are seen to be similar between bright intermediate state transition of different luminosities, other parameters like $T_{\rm in}$, $kT_{\rm e}$ and reflection strength ($R_{\rm s}$) are seen to be sensitive to the luminosity of state transition (\S\ref{subsubsec:2002--2003} \& \S\ref{subsubsec:2004--2005}).

\item In addition to reflection modeling, we also perform a continuum fitting analysis of several spectra across the bright-intermediate state transition of the 2002--2003 outburst, by employing a self-consistent Comptonized accretion disk model accounting for the scatter of disk photons by the corona (\S\ref{subsec:comprehensive}). The mild disk truncation ($\lesssim5.5~R_{\rm g}$) obtained via reflection fitting is corroborated with the $R_{\rm in}$ measurements from continuum fitting analysis.

\item With measured values of the coronal optical depth and $\Gamma$ from the fitting procedure, we solve eqs. \ref{eq:Gamma1} and \ref{eq:Gamma2} to find a strong anti-correlation of spectral hardness with $kT_{\rm e}$. With known values of \texttt{diskbb} norm and estimates of spectral hardening factor ($f_{col}$), we solve eq. \ref{eq:norm} to determine $R_{\rm in}$, and independently find it corroborating with our results from reflection fitting (\S \ref{subsubsec:Tin}).

\item With the inner disk not seen to be physically moving towards the black hole significantly during the bright hard to soft state transition, the changes seen in the model parameters can be attributed to processes like, increase in spectral hardening factor, compactification of corona, collapse of the magnetically supported disk from geometrically thick to thin, increase in $\dot{M}$, among others. The question of the physical mechanism triggering the state transitions still remains uncertain (\S\ref{sec:discussion}).\\

\end{enumerate}

\section*{Acknowledgement} \label{sec:acknowledgement}

N.S. acknowledges the support from DST-INSPIRE, Caltech SURF, and Columbia University Dean's fellowships. J.A.G. acknowledges the support from NASA grant 80NSSC19K1020, and from the Alexander von Humboldt Foundation. R.M.T.C. has been supported by NASA ADAP grant 80NSSC17K0515. V.G. is supported through the Margarete von Wrangell fellowship of the ESF and the Ministry of Science, Research and the Arts Baden-W\"{u}rttemberg. We also thank Ronald A. Remillard for his contribution towards processing the data, which have been used in our analysis. This work was partially supported under NASA contract No. NNG08FD60C.\\

\textit{Facility:} \textit{RXTE} \citep[PCA;][]{1996SPIE.2808...59J}

\textit{Softwares:} \textsc{xspec} v12.9.1q \citep{1996ASPC..101...17A}, \textsc{relxill} v1.4 \citep{2014ApJ...782...76G, 2014MNRAS.444L.100D}, \textsc{xillver} \citep{2010ApJ...718..695G, 2013ApJ...768..146G}, \textsc{pcacorr} \citep{2014ApJ...794...73G}

\textit{Version note:} The erratum in \cite{Sridhar_20} is amended in this document.

\appendix

\section{Potential of \textit{RXTE}/PCA and \textsc{relxill}}\label{appendix:rxte}

In this section, we test the capability of \textit{RXTE}/PCA and \textsc{relxill}, in detecting large values of $R_\mathrm{in}$ for an axisymmetric, stationary source irradiating isotropically \footnote{A similar analysis with different set of parameters of \textsc{relxill} has been performed by \cite{2017ApJ...851...57C}.}. Determining the capability of \textit{RXTE}/PCA and \textsc{relxill} to detect such narrow features is essential to conclude whether the inner disk is actually truncated at $R_\mathrm{in}>>R_\mathrm{ISCO}$ or not. For this exercise, we generate synthetic spectra using \texttt{fakeit} routine in \textsc{xspec v12.9.1q} \citep{1996ASPC..101...17A}. The simulated spectra are created with the response of \textit{RXTE}/PCA in the $\sim$3--45 keV energy band with long enough exposure such that they have at least a million counts in the considered energy range. 

\begin{figure*}[ht] \centering
\includegraphics[width=1\linewidth]{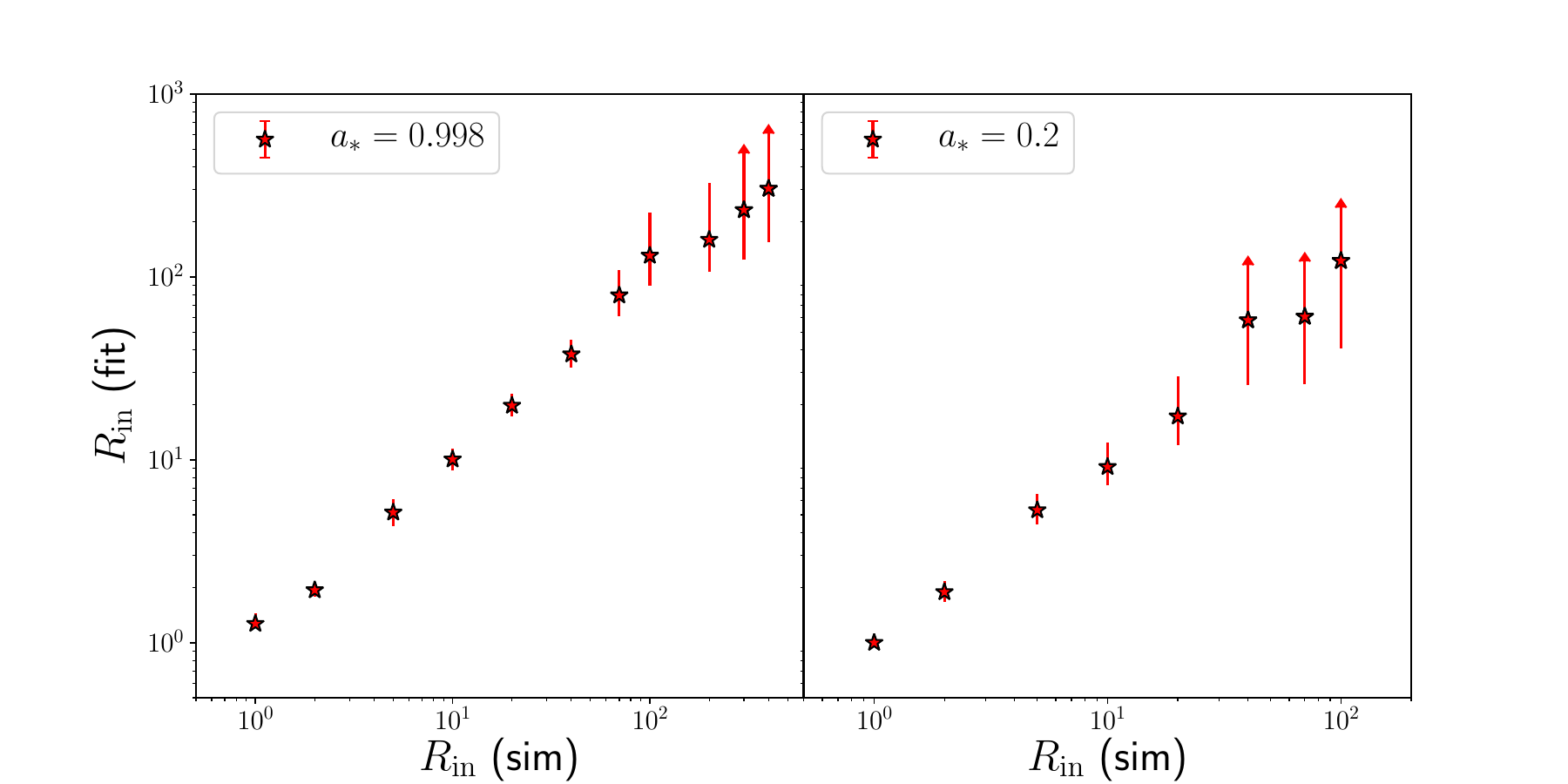}
\caption{Best fit $R_\mathrm{in}$ values (vertical axes; in units of $R_\mathrm{ISCO}$) plotted against the $R_{\rm in}$ used to simulate the spectra (horizontal axes). The error bars represent 90\% confidence intervals. For $a_{\star}$=0.998, the fit goes unconstrained for a value of $R_\mathrm{in}\gtrsim$ 120 $R_\mathrm{ISCO}$. On the other hand, for a small spin parameter of $a_{\star}\sim$ 0.2, the fit goes unconstrained for an $R_\mathrm{in}$ value as low as $\sim$40 $R_\mathrm{ISCO}$. The parameters fed-in to simulate the spectra in the left and right panels are described in Appendix \ref{appendix:rxte}.}
\label{fig:rxte_capability}
\end{figure*}

Using these specifications, 1000 spectra are synthesized with model: \texttt{TBabs*relxill}, with varying $R_\mathrm{in}$ values for two cases: (i) $R_\mathrm{f}=0.2$, $a_{\ast}=0.998$ and (ii) $R_\mathrm{f}=0.7$, $a_{\ast}=0.2$, where $R_\mathrm{f}$ is the reflection fraction and $a_{\ast}$ is the dimensionless spin parameter. In both the cases, the disk emissivity is fixed to the canonical value of 3 \citep{1989MNRAS.238..729F}, inclination to $45^ \circ$, photon index $\Gamma$ to 2, log$(\xi)$ to 3 and the iron abundance $A_\mathrm{Fe}$ to 5 ($\times$Solar). The input values are chosen such that they represent the typical physical conditions present in most observed systems \citep{2013ApJ...768..146G}.
The simulated spectra are then fitted with the model: \texttt{TBabs*relxill}, where all the parameters are frozen to simulated values with $R_\mathrm{in}$ as a free parameter. For cases (i) and (ii), $R_\mathrm{in}$ cannot be constrained for values $\gtrsim 120~R_\mathrm{ISCO}$, and $\gtrsim 40~R_\mathrm{ISCO}$ respectively (Figure \ref{fig:rxte_capability}). This result shows that for a high-spin black hole source like that of GX 339--4, the data of \textit{RXTE}/PCA with the \textsc{relxill} model is capable of constraining the inner radius of the accretion disk as large as $120~R_\mathrm{ISCO}$.

\section{Disparity with the truncated disk model}\label{appendix:truncation}

Having demonstrated the capability of {\it RXTE}/PCA and \textsc{relxill} in detecting large disk truncation radius (Appendix \ref{appendix:rxte}), we now simulate spectra corresponding to large disk truncation with the PCA instrument response. We use the best fit model parameters (Model 2(i)) of \cite{2016MNRAS.458.2199B} to simulate the spectra of their observations 3 and 7 using PCA's response. The parameter values used for simulating the spectra are listed in Table \ref{tab:Basak_table}. \cite{2016MNRAS.458.2199B} had analyzed seven observations of GX 339--4 in its hard state, as observed by \textit{XMM-Newton} EPIC-pn detector. For our exercise, we choose the timing mode observations, 3 and 7 from their list because of their high count rates, and the `timing mode' of the observation. In addition, we simulate another spectrum with all the same parameters from fitting their observation 7, but with $R_\mathrm{in}=R_\mathrm{ISCO}$, for comparison purpose.

In order to visualize the Fe K reflection feature, we fit the simulated spectra with a simple absorbed power-law model, and follow the procedures outlined in \S \ref{subsec:ratio}. Broadening of the Fe K line with decreasing $R_{\rm in}$ values is seen in Figure \ref{fig:Basak_ratio}, as expected. Note that observations 7 and 3 correspond to the bright softest and hardest state respectively, in the sample of \cite{2016MNRAS.458.2199B}. Should the disk indeed be truncated at $\sim100~R_{\rm ISCO}$ during any of our selected bright hard to soft state observations, we would expect to see broadened emission feature as seen in Figure \ref{fig:Basak_ratio}. However, our result (see Figure \ref{fig:Fe_ratio}) does not exhibit any line broadening, indicative of a stable inner disk radius during the bright hard to soft state transition spanning across a wide range of hardness values. As discussed in \cite{2010ApJ...724.1441M}, this disparity can be attributed to the deteriorating effects of high photon pile-up in the timing mode observations of \textit{XMM-Newton}. If not corrected for the pile-up effects, it can artificially make the continuum softer, resulting in a narrower Fe K profile, thereby leading to estimates of truncation much larger than reality.

  \begin{minipage}{\textwidth}
  \begin{minipage}[b]{0.49\textwidth}
    \centering
	\includegraphics[width=1\linewidth]{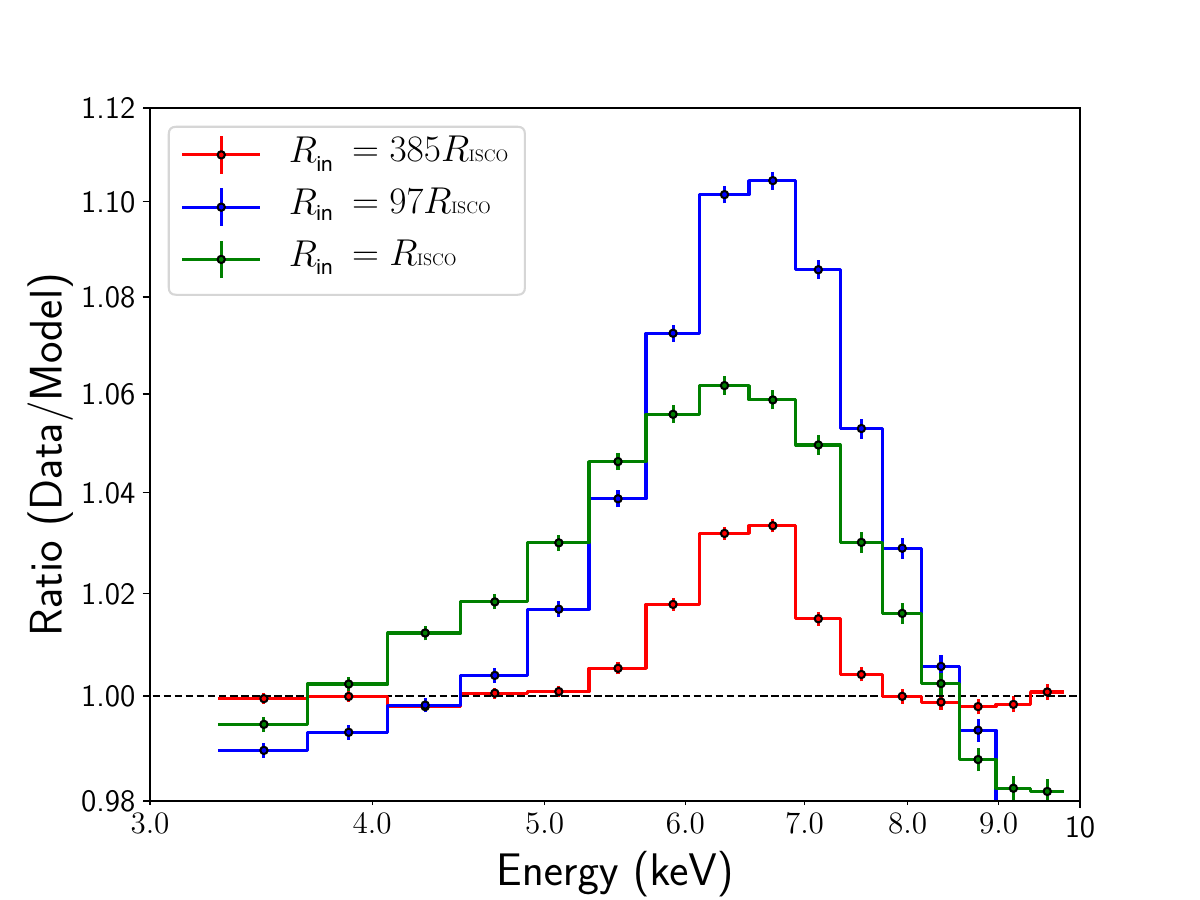}
        \figcaption{Data/model ratio plot produced by fitting the simulated spectrum from \cite{2016MNRAS.458.2199B} (see Table \ref{tab:Basak_table}) using PCA's response with absorbed \texttt{powerlaw} model. Red and blue profiles correspond to their observations 3 and 7 respectively. Green profile is simulated for the same set of parameters except for the disk truncation fixed at $R_\mathrm{ISCO}$, for comparison purpose.} \label{fig:Basak_ratio}
        
  \end{minipage}
  \hfill
  \begin{minipage}[b]{0.49\textwidth}
    \centering
\tabcaption{Parameter values from fitting the best fit model 2(i) of \cite{2016MNRAS.458.2199B} to the data of their observations 3 and 7, used for simulating the spectra with PCA's response.} \label{tab:Basak_table}
\begin{tabular}{cccc} 
  \hline
  \hline
  Component & Parameter & Obs. 3 & Obs. 7\\
  \hline
  \texttt{TBabs} & $N_{\rm H}$ [cm$^{-2}$]& \multicolumn{2}{c}{0.71}\\
  \texttt{nthComp} & $kT_\mathrm{e}$ [keV]& \multicolumn{2}{c}{100}\\  
  \texttt{nthComp} & $i$ [deg] & \multicolumn{2}{c}{$43.8$}\\  
  \texttt{relxill} & $A_\mathrm{Fe}$ [Solar] & \multicolumn{2}{c}{1}\\
  \texttt{relxill} & $a_{\ast}$ & \multicolumn{2}{c}{0.998}\\
  \texttt{relxill} & $E_\mathrm{cut}$ [keV]& \multicolumn{2}{c}{150}\\  
  \texttt{relxill} & $N_{\rm rel}$ & \multicolumn{2}{c}{0.04}\\
  \hline    
  \texttt{diskbb} & $T_\mathrm{in}$ [keV]& 0.19 & 0.21\\
  \texttt{diskbb} & $N_{\rm disk}$ & $7.74 \times 10^5$ & $4.95 \times 10^4$\\
  \texttt{nthComp} & $\Gamma$ & 1.56 & 1.69\\
  \texttt{nthComp} & $N_{\rm nthC}$ & 0.19 & 0.52\\
  \texttt{relxill} & $R_\mathrm{in} [R_\mathrm{ISCO}]$ & 385 & 97\\
  \texttt{relxill} & log($\xi$) [erg cm s$^{-1}$] & 2.69 & 3.31\\
  \hline
\end{tabular}
\tablecomments{A short description of each parameter is provided under Table \ref{tab:2002--2003}.}
  \end{minipage}
  \end{minipage}

\section{MCMC parameter probability distributions}\label{appendix:mcmc}

\begin{figure*}[ht]
\begin{minipage}[c]{\textwidth}
\gridline{\fig{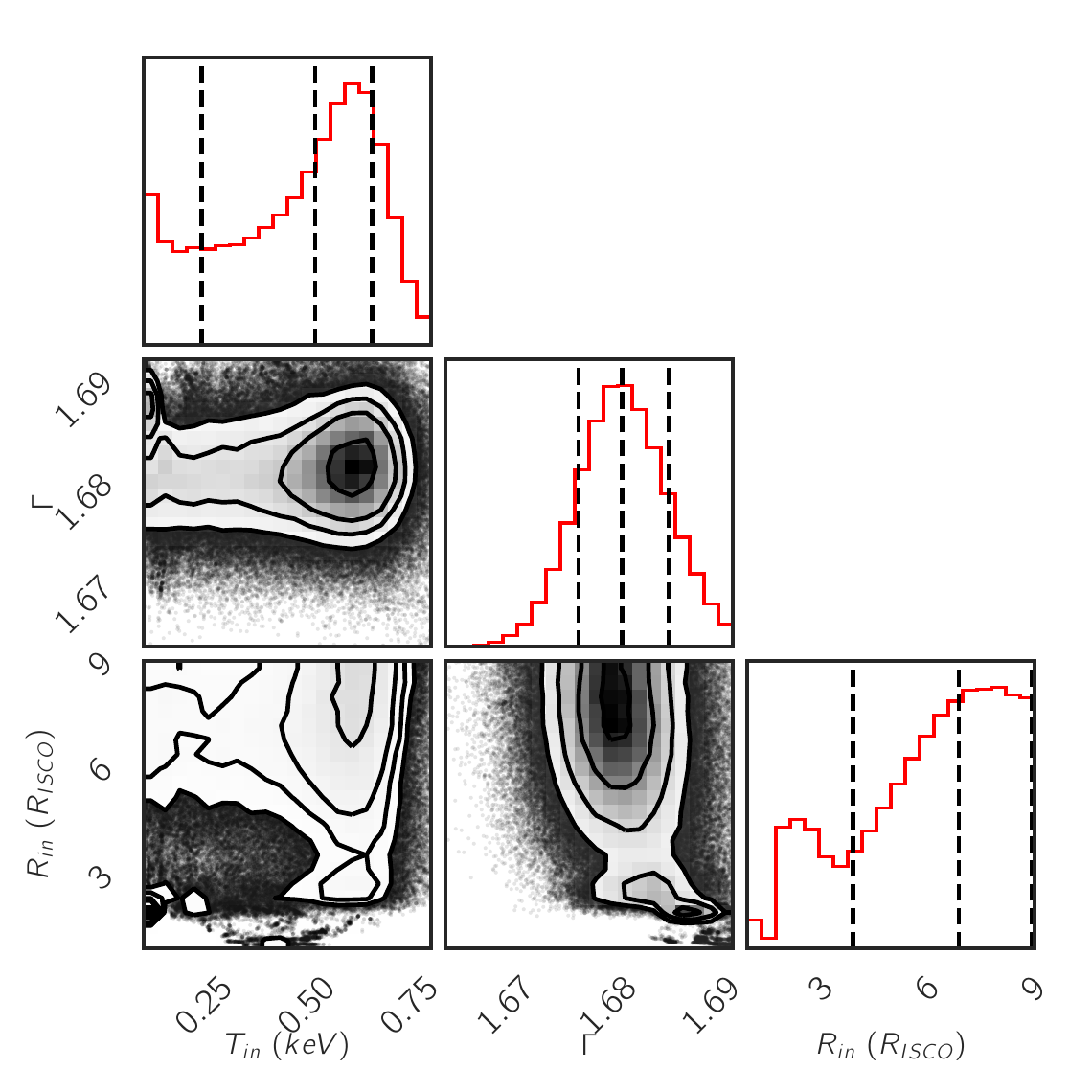}{0.35\textwidth}{(a)$HR\simeq0.80$}
			\fig{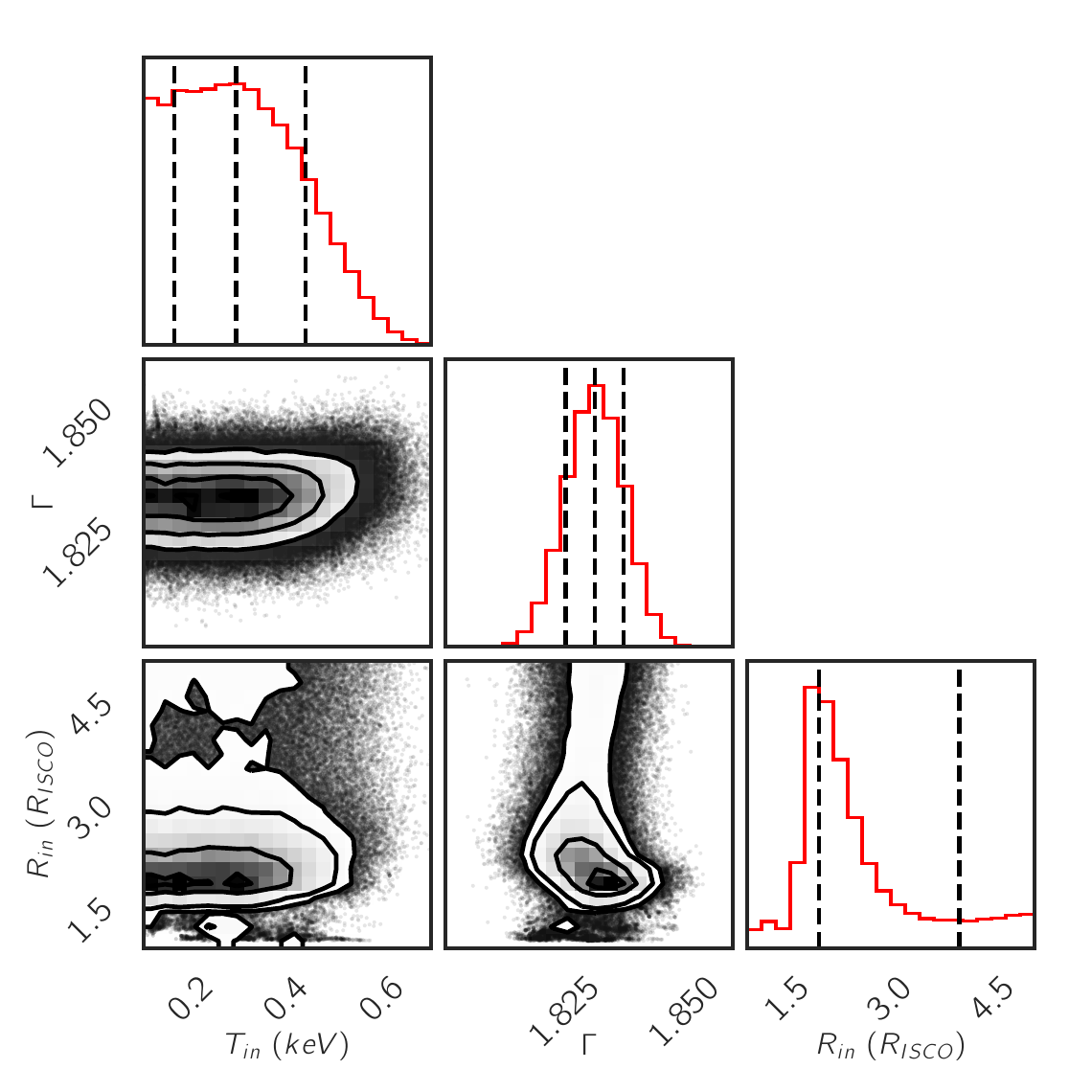}{0.35\textwidth}{(b)$HR\simeq 0.70$}
			\fig{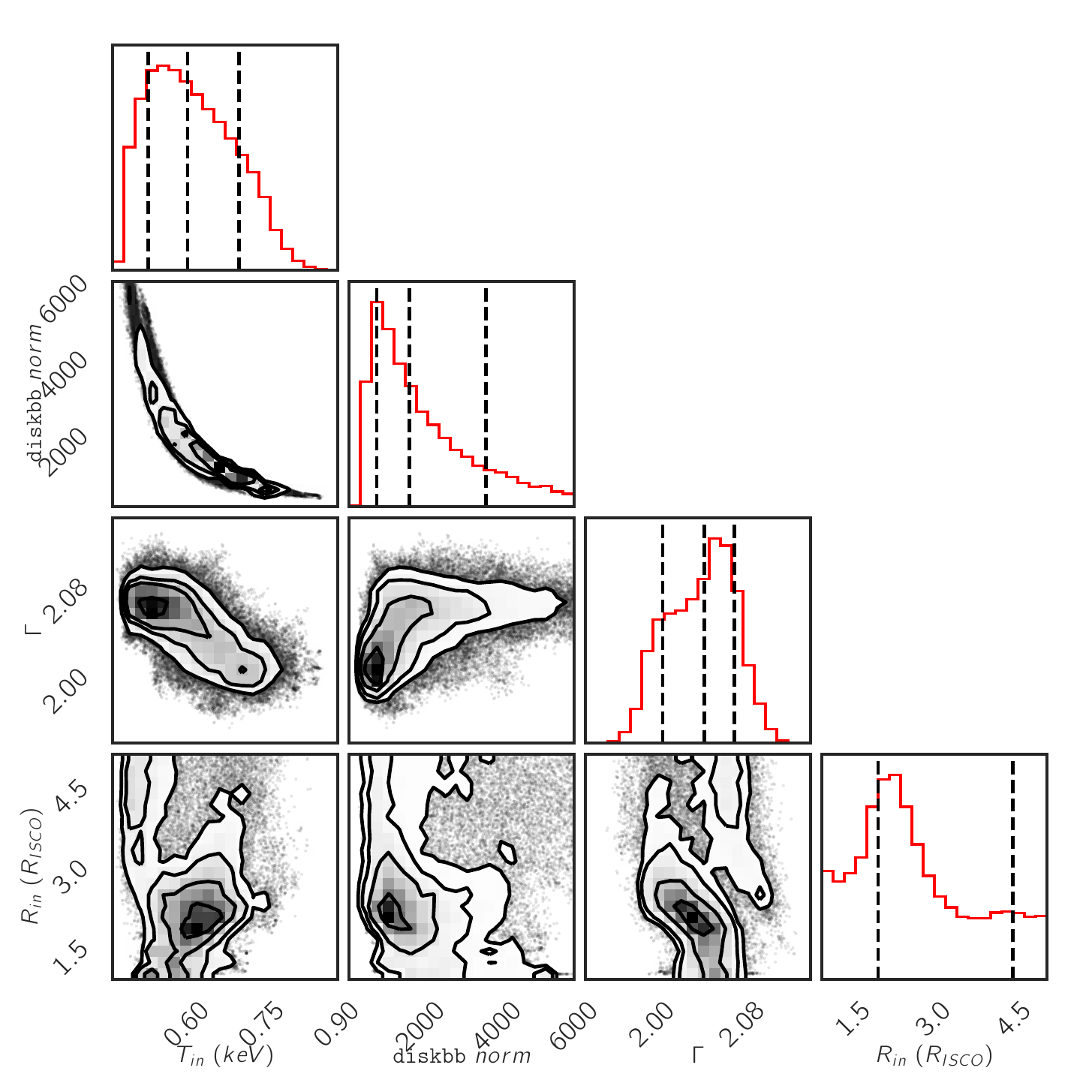}{0.35\textwidth}{(c)$HR\simeq 0.56$}}
\gridline{\fig{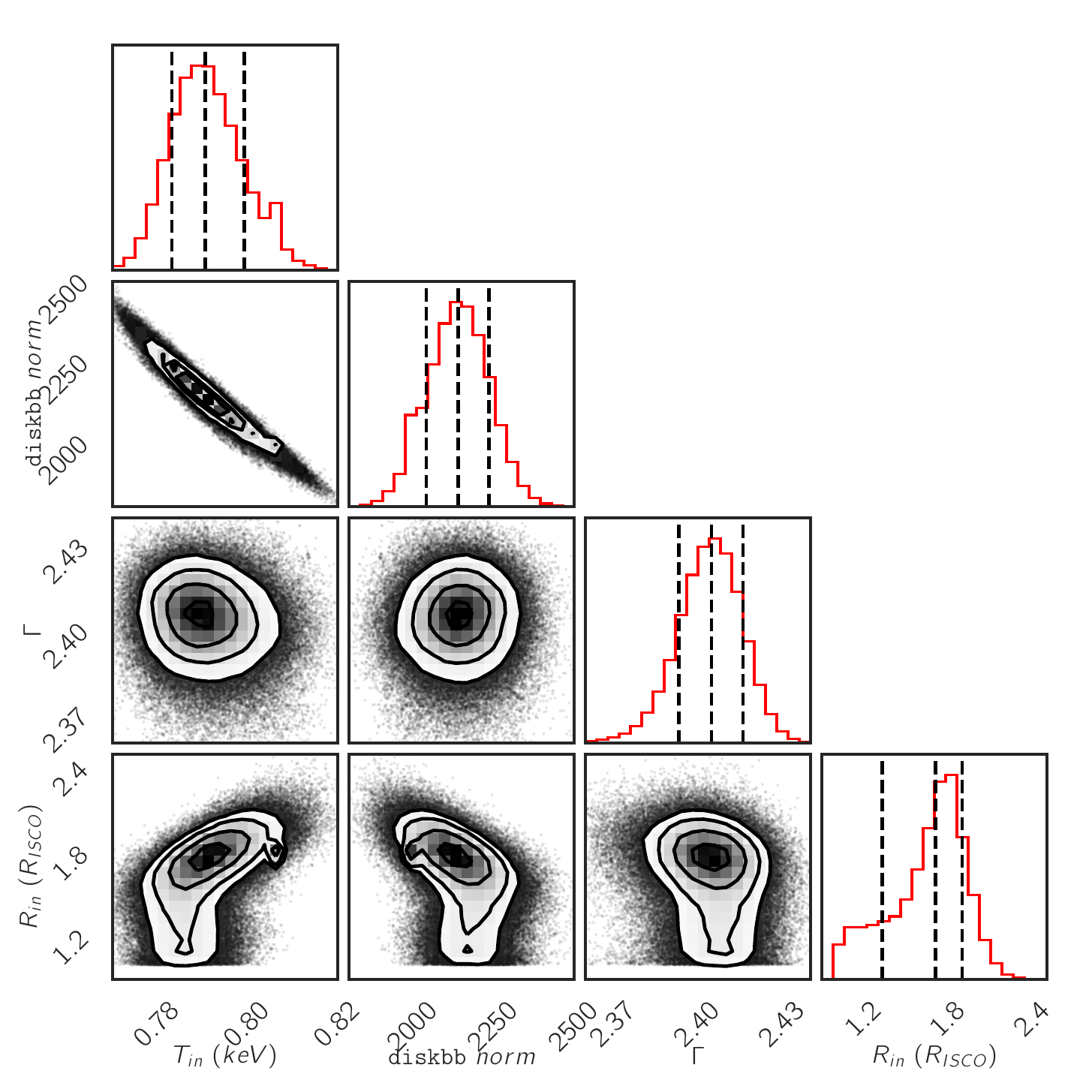}{0.35\textwidth}{(d)$HR\simeq 0.40$}
			\fig{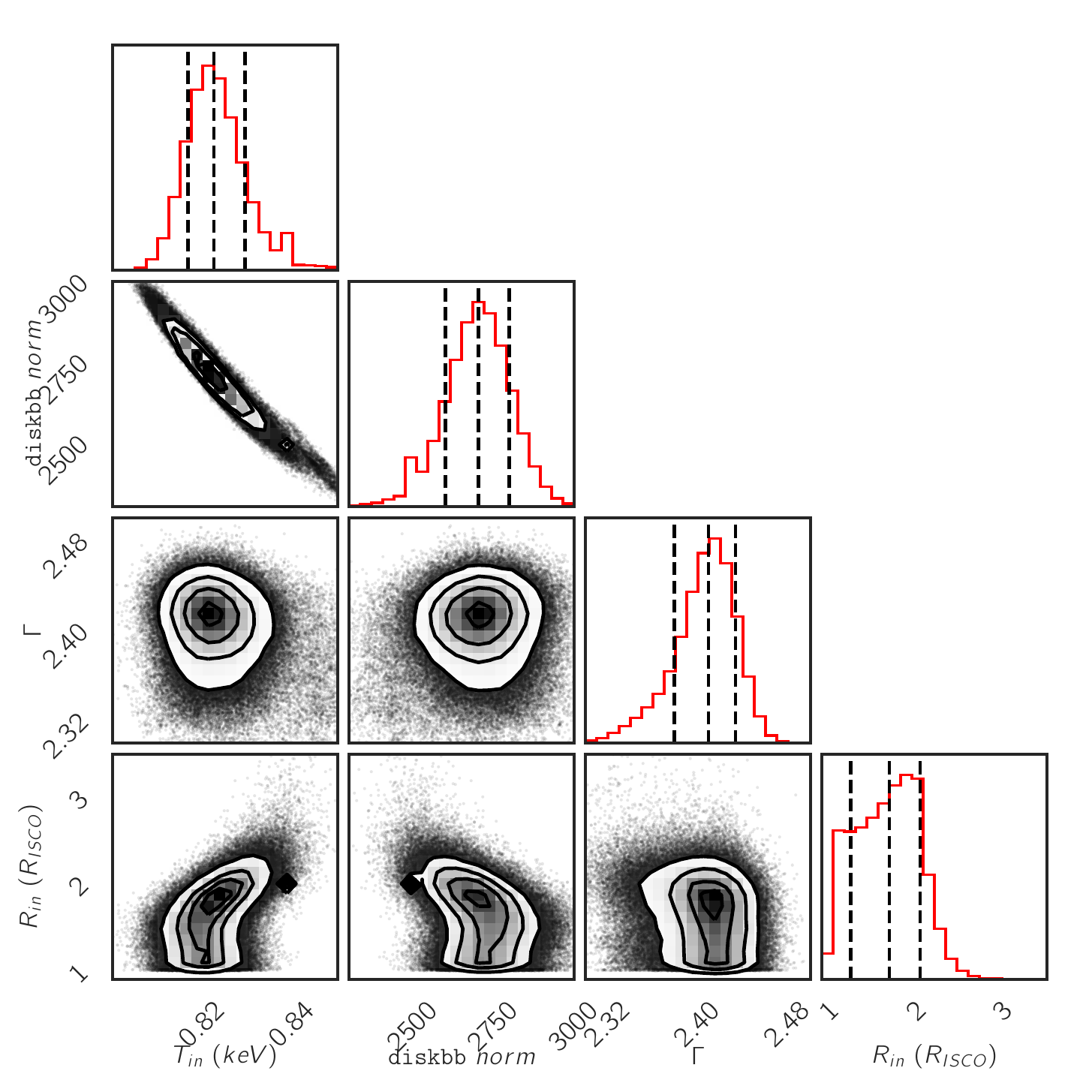}{0.35\textwidth}{(e)$HR\simeq 0.30$}
			\fig{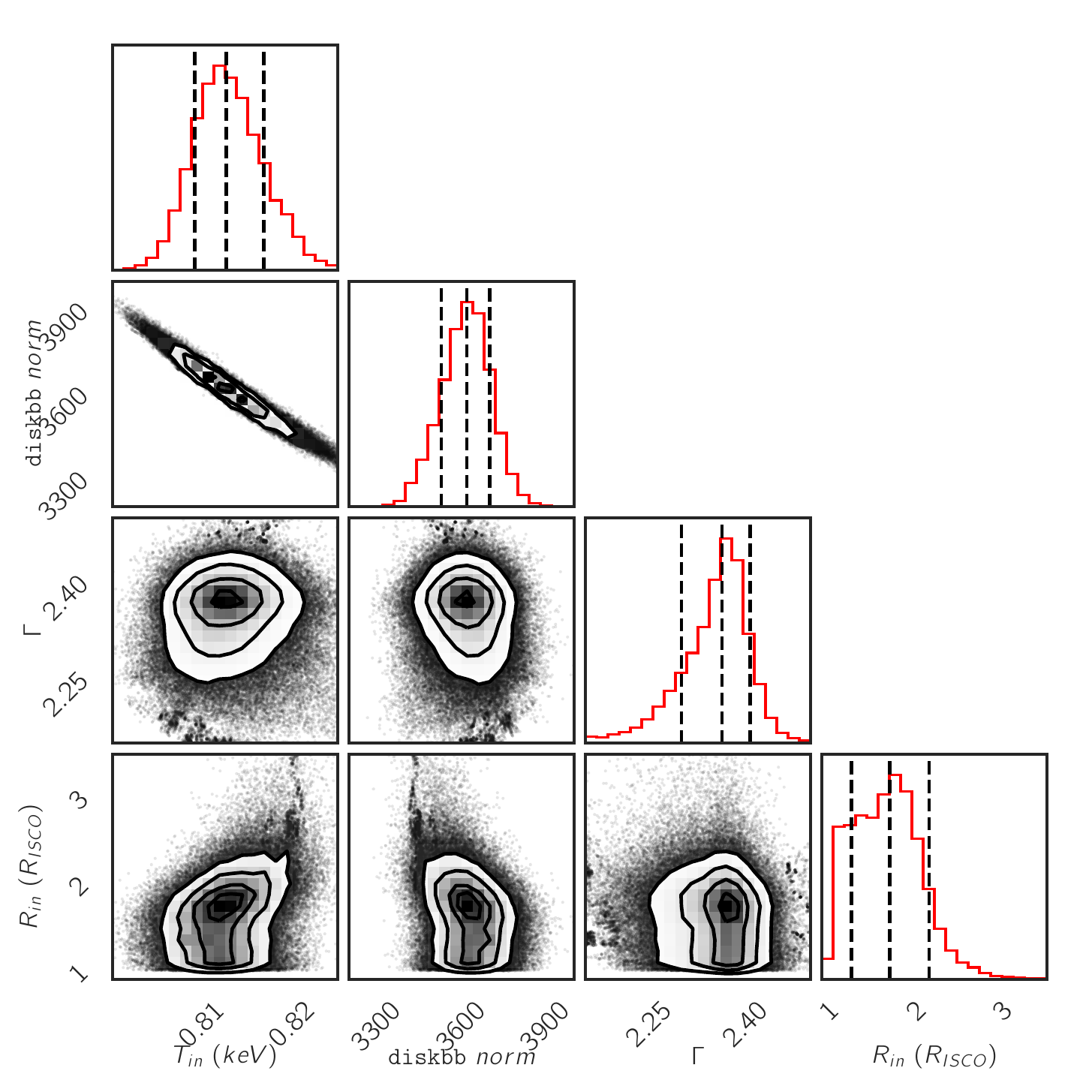}{0.35\textwidth}{(f)$HR\simeq 0.20$}}
\gridline{\fig{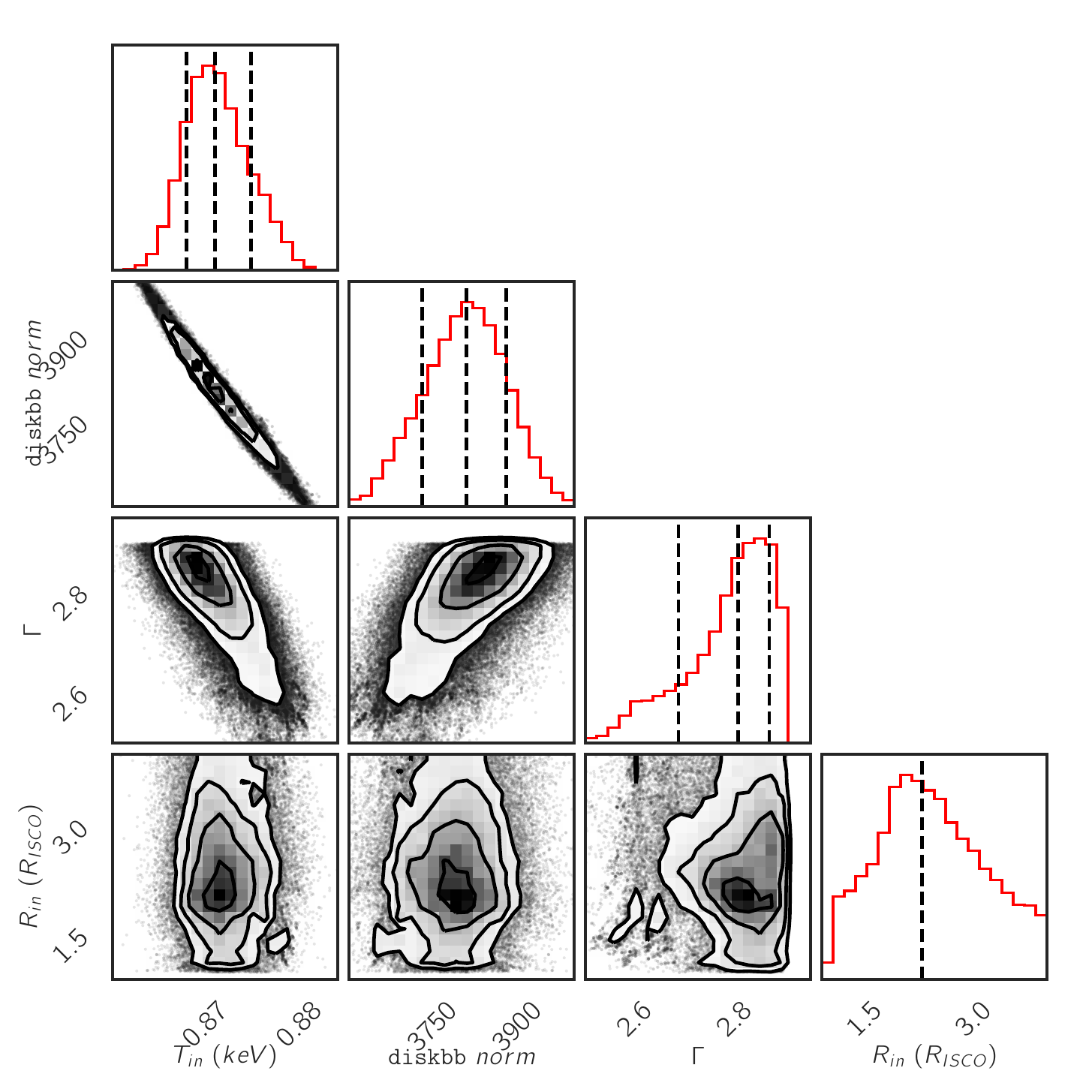}{0.35\textwidth}{(g)$HR\simeq 0.10$}}
\caption{One and two dimensional projections of the posterior probability distributions, and the 0.16, 0.5 and 0.84 quantile contours derived from the MCMC analysis for the inner disk temperature ($T_\mathrm{in}$), \texttt{diskbb} norm, photon index ($\Gamma$) and the inner truncation radius ($R_{\rm in}$), measured using reflection spectroscopy for different spectral hardness values across the bright-intermediate states of the 2002--2003 outburst. Note that the value of \texttt{diskbb} norm is frozen to $0$ for $HR\simeq0.8$ and $HR\simeq0.7$. These figures are produced using the {\tt corner} package \citep{2016JOSS....1...24F}.}
\label{fig:corner_2002--2003}
\end{minipage}
\end{figure*}

\begin{figure*}[ht]
\begin{minipage}[c]{\textwidth}
\gridline{\fig{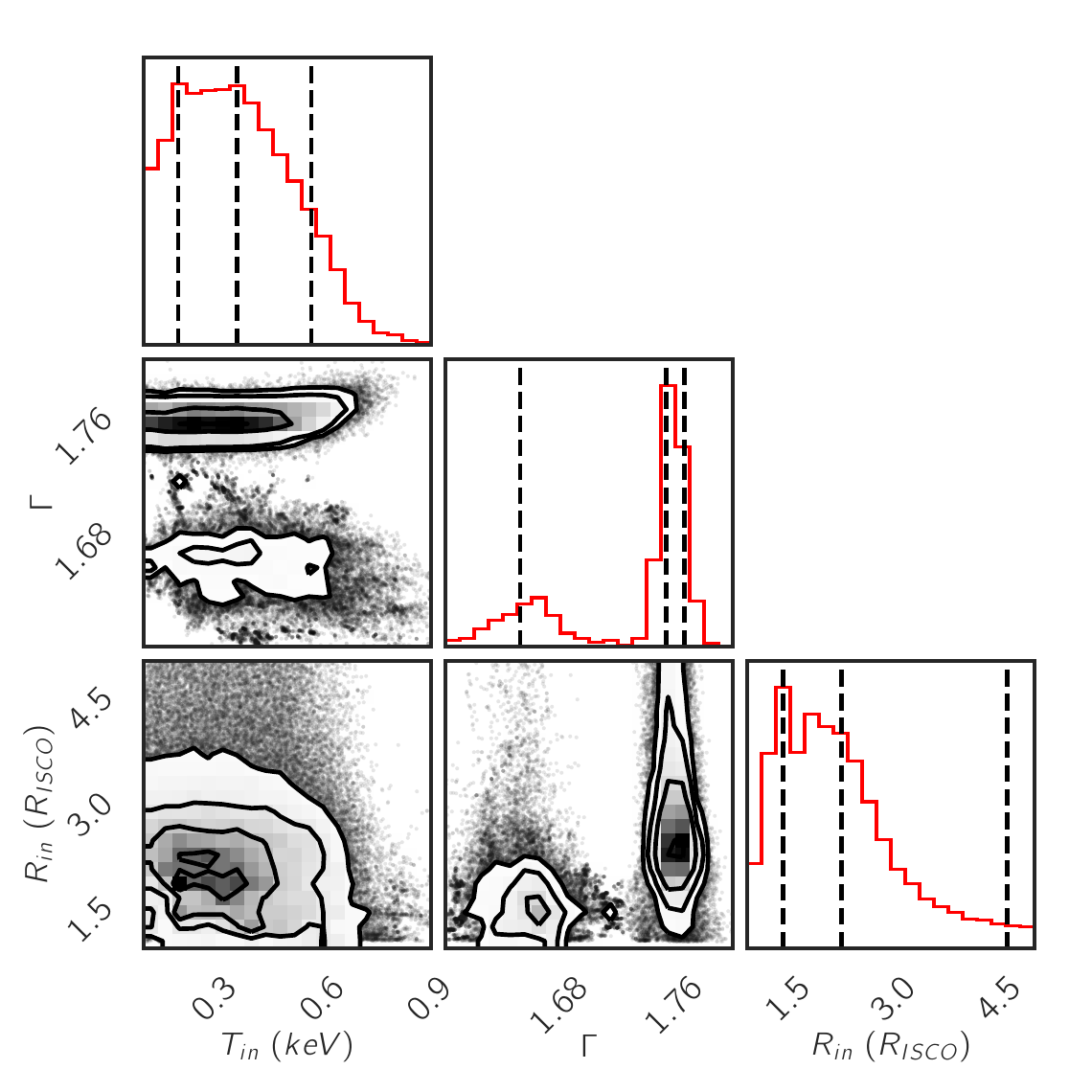}{0.35\textwidth}{(a)$HR\simeq 0.80$}
			\fig{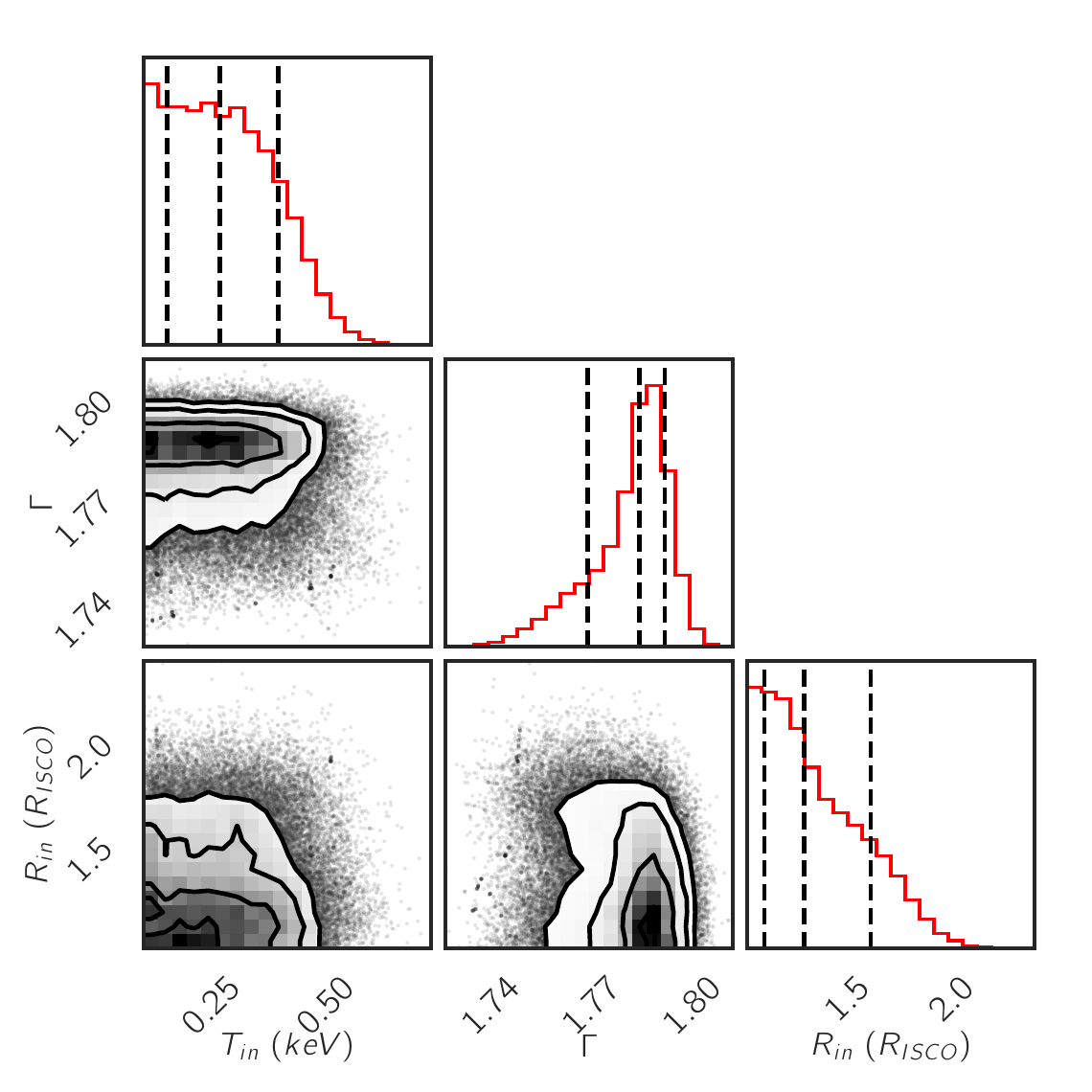}{0.35\textwidth}{(b)$HR\simeq 0.72$}
			\fig{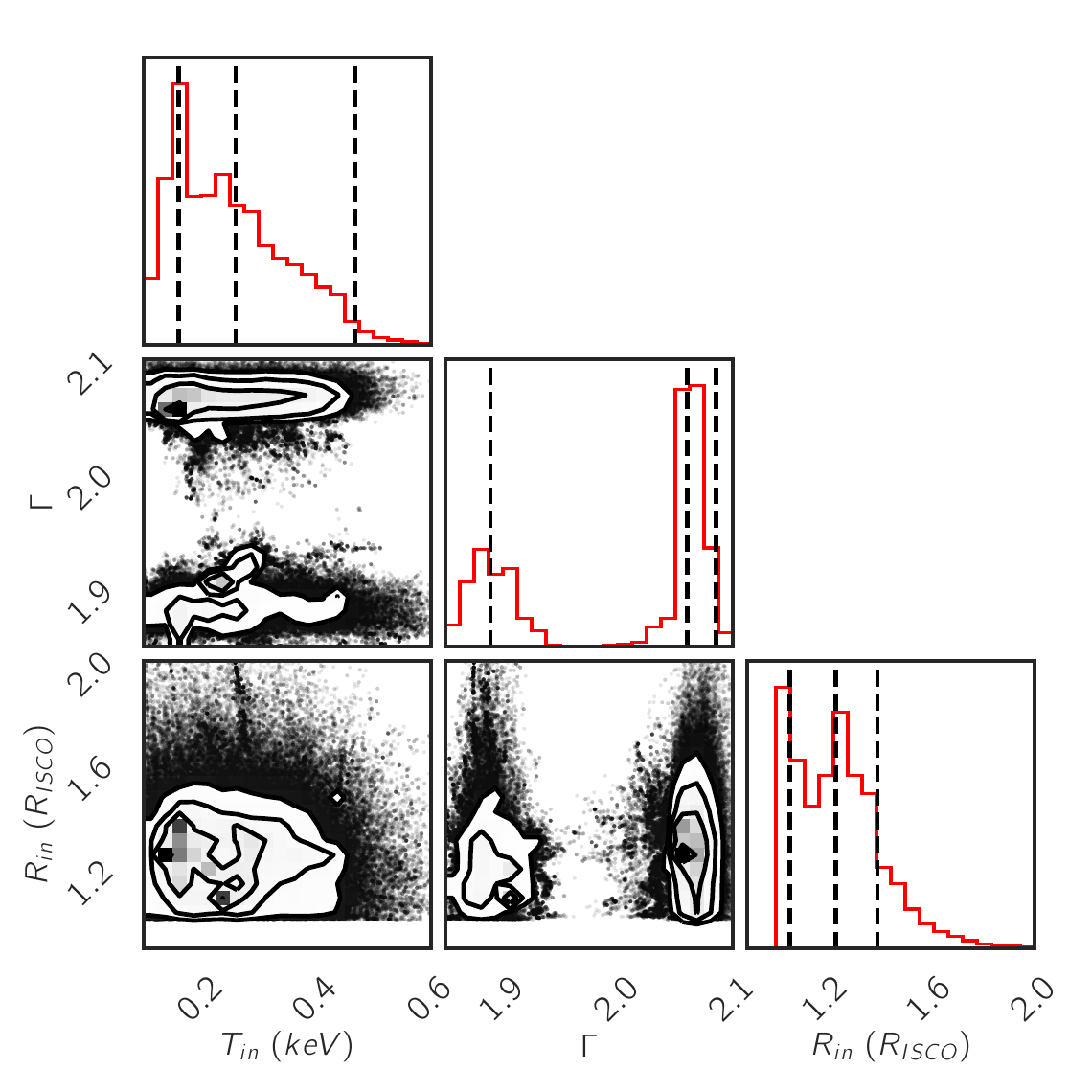}{0.35\textwidth}{(c)$HR\simeq 0.65$}}
\gridline{\fig{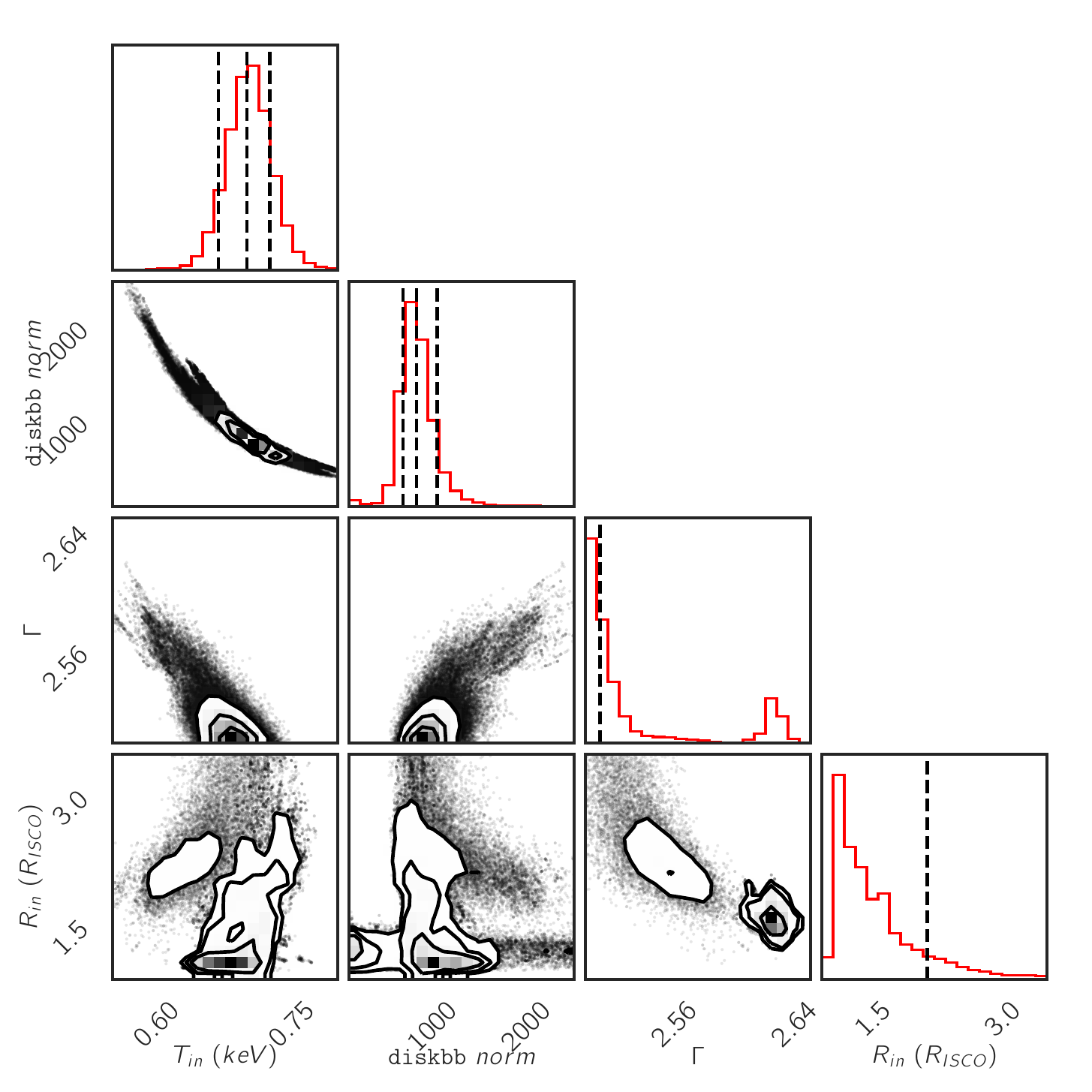}{0.35\textwidth}{(d)$HR\simeq 0.48$}
			\fig{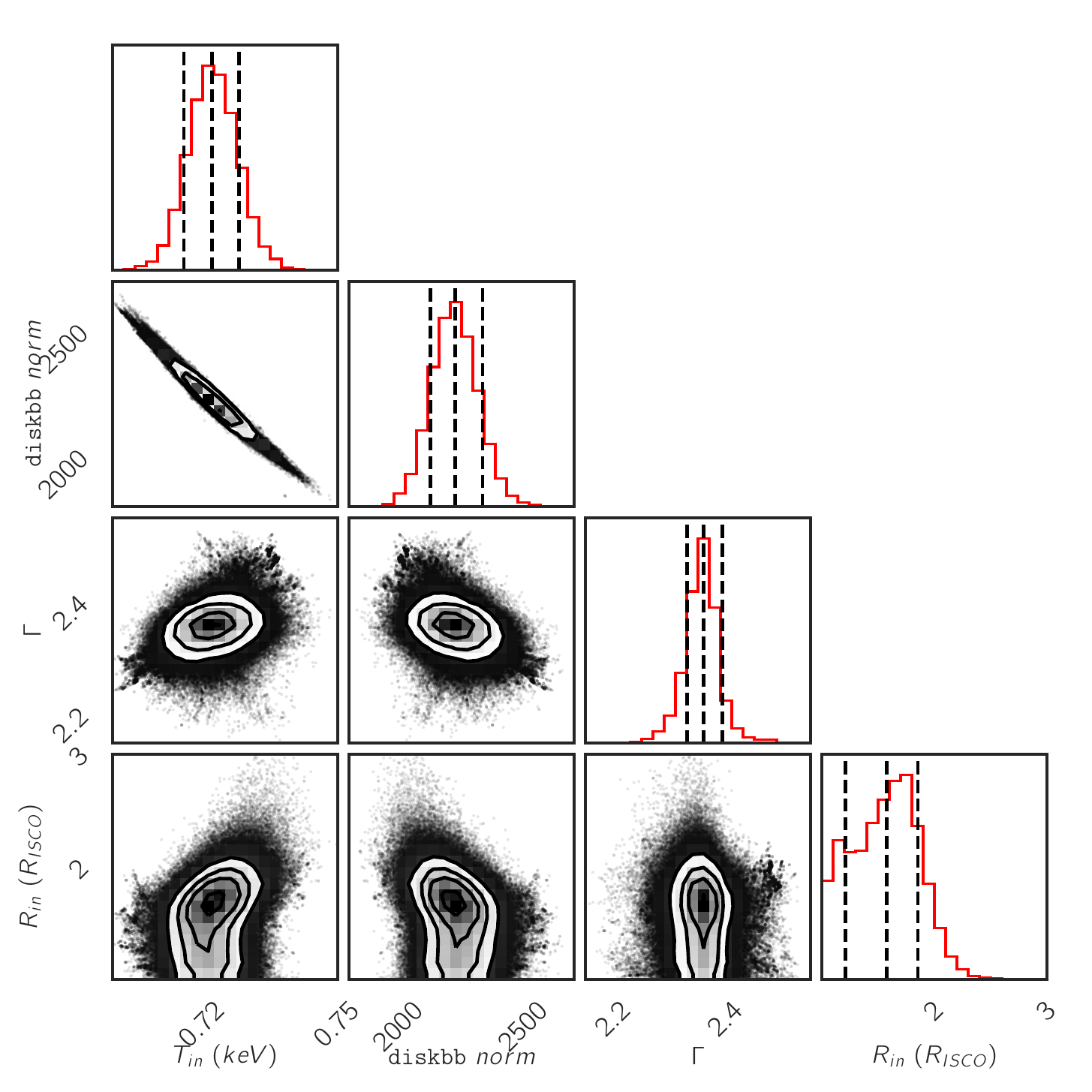}{0.35\textwidth}{(e)$HR\simeq 0.30$}
			\fig{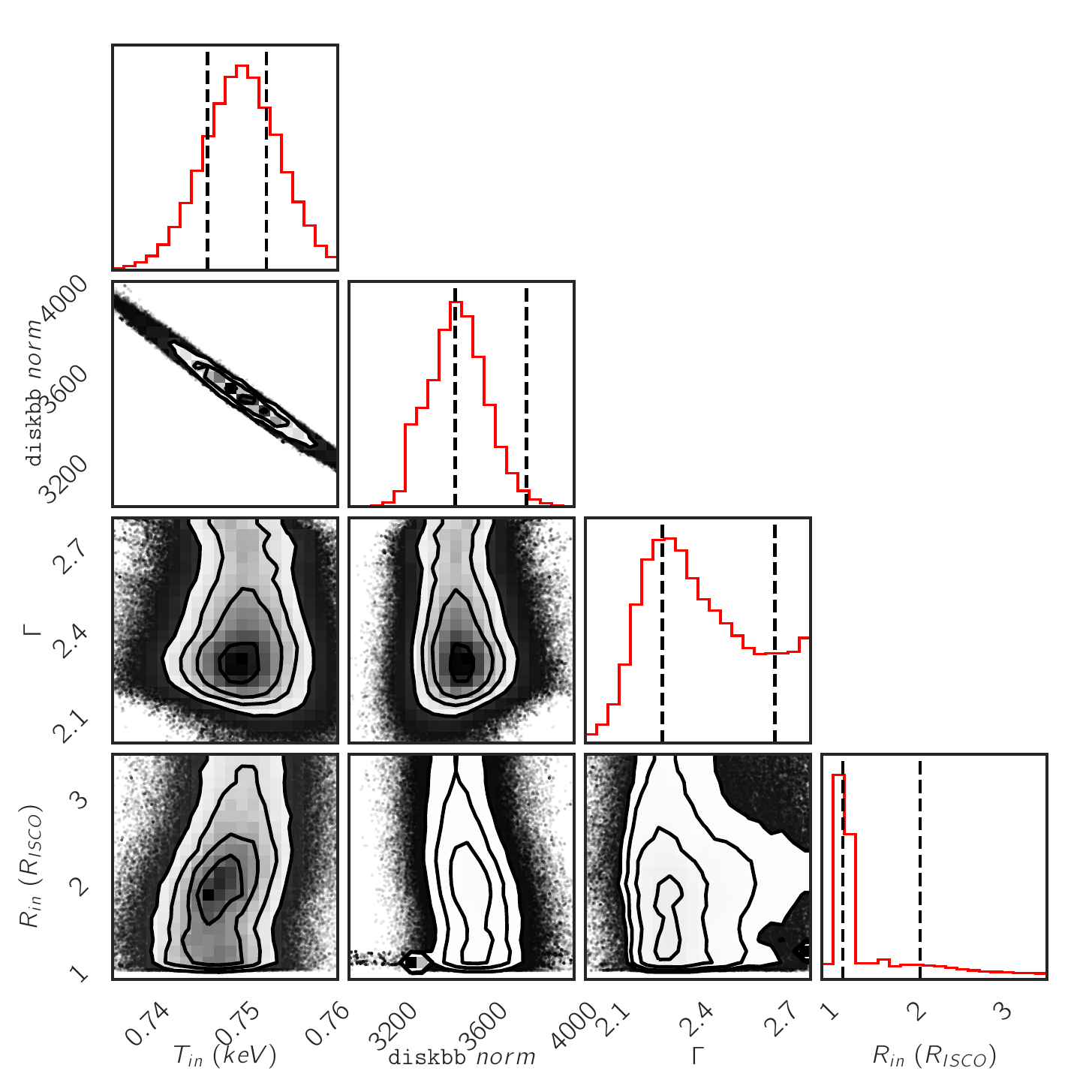}{0.35\textwidth}{(f)$HR\simeq 0.13$}}
\gridline{\fig{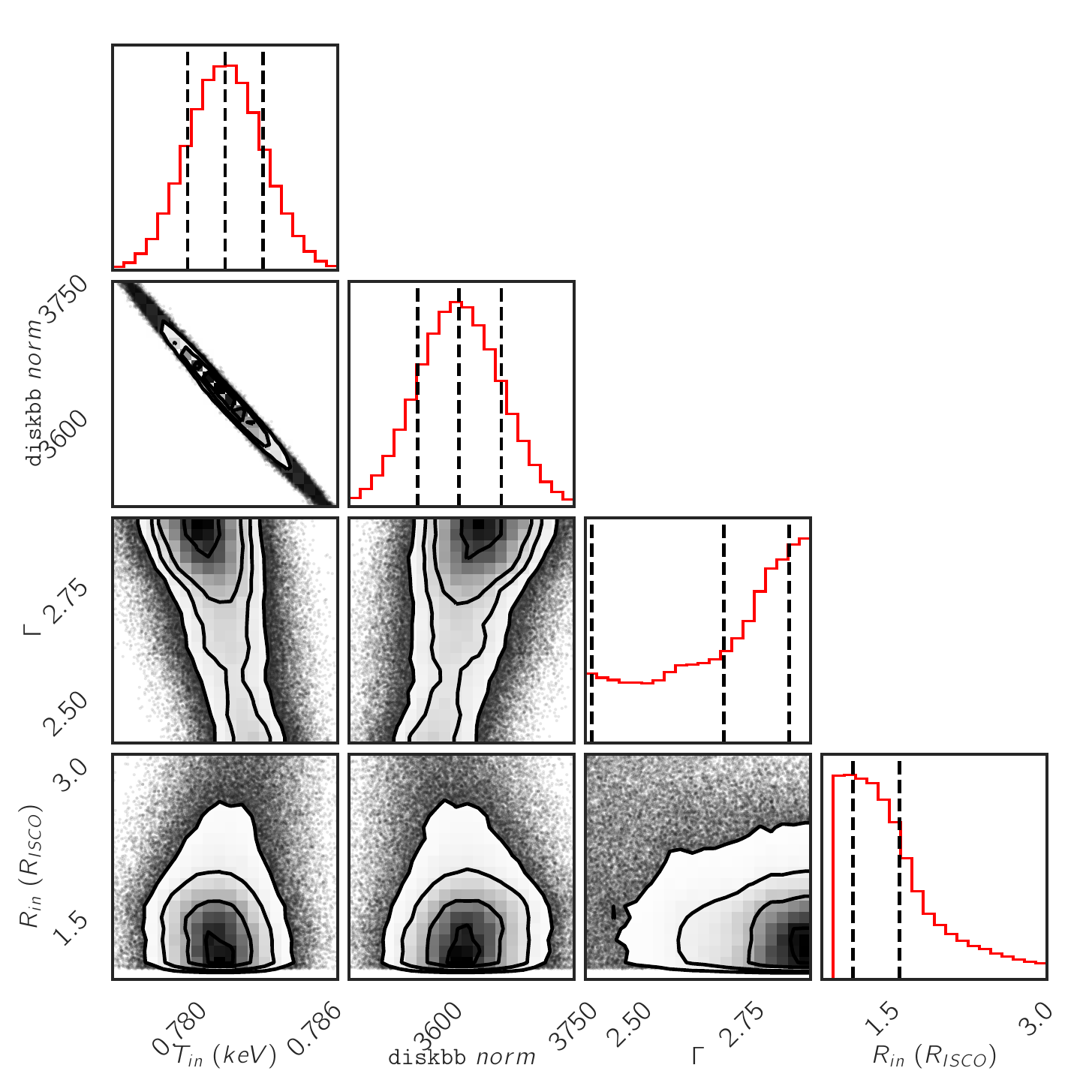}{0.35\textwidth}{(g)$HR\simeq 0.05$}}

\caption{One and two dimensional projections of the posterior probability distributions, and the 0.16, 0.5 and 0.84 quantile contours derived from the MCMC analysis for the inner disk temperature ($T_\mathrm{in}$), \texttt{diskbb} norm, photon index ($\Gamma$) and the inner truncation radius ($R_{\rm in}$), measured using reflection spectroscopy for different spectral hardness values across the bright-intermediate states of the 2004--2005 outburst. Note that the value of \texttt{diskbb} norm is frozen to $0$ for $HR\simeq0.8$, $HR\simeq0.72$ and $HR\simeq0.65$. These figures are produced using the {\tt corner} package \citep{2016JOSS....1...24F}.}
\label{fig:corner_2004--2005}

\end{minipage}
\end{figure*}

In this section, we show the contour maps and probability distributions for the set of most relevant physical parameters derived using the MCMC analysis (Figures \ref{fig:corner_2002--2003} \& \ref{fig:corner_2004--2005}). These contour maps demonstrate how well each parameter is constrained in the fit, and the level of correlation and degeneracy between parameters. For each map, we also show the 0.16, 0.5, and 0.84 quantiles. Refer to \S\ref{subsec:MCMC} for a detailed description of the MCMC setup and the results.

\bibliographystyle{yahapj}
\bibliography{main}

\end{document}